\documentclass[ba]{imsart}
\pubyear{2021}
\volume{TBA}
\issue{TBA}
\firstpage{1}
\lastpage{1}
\usepackage[sectionbib]{natbib}
\usepackage{array,fancyheadings,rotating,supertabular}
\usepackage[colorlinks,citecolor=blue,urlcolor=blue,filecolor=blue,backref=page]{hyperref}
\usepackage{xcolor}
\usepackage{amsfonts}
\usepackage{multirow}

\usepackage{sectsty, secdot}
\sectionfont{\fontsize{12}{14pt plus.8pt minus .6pt}\selectfont}
\renewcommand{\theequation}{\thesection\arabic{equation}}
\subsectionfont{\fontsize{12}{14pt plus.8pt minus .6pt}\selectfont}

\usepackage{times}
\usepackage{bm,amsbsy,amsmath,amsthm,amssymb,times,graphicx,enumerate,color,subfigure,psfrag} 
\usepackage[ruled,vlined]{algorithm2e}
\usepackage{enumitem}
\usepackage{supertabular}

\newcommand{\comments}[1]{}
\newcommand{\pcite}[1]{\citeauthor{#1}'s \citeyearpar{#1}}

\newtheorem{corollary}{Corollary}
\newtheorem{lemma}{Lemma}
\newtheorem{definition}{Definition}
\newtheorem{theorem}{Theorem}
\newtheorem{remark}{Remark}

\newcommand{\Real}{\mathbb{R}}

\newcommand{\ind}{\buildrel \text{ind} \over \sim}

\DeclareMathOperator*{\argmax}{argmax}
\DeclareMathOperator*{\argmin}{argmin}

\def\shalf{{\mbox{{\footnotesize$\frac{1}{2}$}}}}

\def\shalf{\mbox{{\footnotesize$\frac{1}{2}$}}}

\newcommand{\sN}{{\cal N}}

\allowdisplaybreaks

\newcommand{\half}{\dfrac{1}{2}}
\newcommand{\rss}{\mathrm{RSS}}
\newcommand{\rssl}{\rss_\lambda}

\def\baro{\vskip  .2truecm\hfill \hrule height.5pt \vskip  .2truecm}
\def\barba{\vskip -.1truecm\hfill \hrule height.5pt \vskip .4truecm}
\def\shalf{\mbox{{\footnotesize$\frac{1}{2}$}}}

\begin{document}
\begin{frontmatter}
\title{Bayesian iterative screening in ultra-high dimensional linear regressions}
\runtitle{Bayesian iterative screening}

\begin{aug}
\author{\fnms{Run} \snm{Wang}\thanksref{addr1}}, 
\author{\fnms{Somak} \snm{Dutta}\thanksref{addr1}}
\and
\author{\fnms{Vivekananda} \snm{Roy}\thanksref{addr1}}%

\runauthor{Wang Dutta and Roy}

\address[addr1]{Department of Statistics, Iowa
   State University, USA 
}

\end{aug}
\begin{abstract}
    Variable selection in ultra-high dimensional linear regression
  is often preceded by a screening step to significantly reduce the
  dimension. Here we develop a Bayesian variable screening
  method (BITS) guided by the posterior model probabilities. BITS can successfully integrate
  prior knowledge, if any, on effect sizes, and the number of true
  variables. BITS iteratively includes potential
  variables with the highest posterior probability accounting for the
  already selected variables. It is implemented by a fast
  Cholesky update algorithm and is shown to have the screening
  consistency property. BITS is built based on a model with Gaussian errors, yet, the screening consistency is proved to hold  under more general tail conditions. The notion of posterior screening
  consistency allows the resulting model to provide a good starting point for further Bayesian variable selection methods. A new screening consistent stopping rule based on posterior probability is developed. Simulation studies and real data examples are used to
  demonstrate scalability and fine screening performance.

\end{abstract}



\end{frontmatter}

\def\thefigure{\arabic{figure}}
\def\thetable{\arabic{table}}

\renewcommand{\theequation}{\thesection.\arabic{equation}}

\fontsize{12}{14pt plus.8pt minus .6pt}\selectfont

\section{Introduction}
\label{sec:int}
These days, in diverse disciplines, data sets with hundreds of
thousands of variables are commonly arising although only a very few
of these variables are believed to be relevant for the response. Thus
variable selection in linear regression has been a major topic of
research over the last two decades in both frequentist and Bayesian
statistics.  A common approach to variable selection as well as
coefficient estimation is by penalizing a loss function. These
shrinkage methods include, but are not limited to, the ridge
regression \citep{hoer:kenn:1970}, the popular Lasso
\citep{tibs:1996}, the bridge regression \citep{huan:horo:ma:2008},
the SCAD \citep{fan:li:2001}, the elastic net \citep{zou:hast:2005},
the Dantzig selector \citep{cand:tao:2007,mazu:radc:2017} and the
adaptive Lasso \citep{zou:2006}.

Several Bayes and empirical Bayes
penalized regression methods have been developed in the
literature. See, for example, \cite{park:case:2008},
\cite{kyun:gill:ghos:case:2010}, and
 \cite{roy:chak:2017}.
Another approach to Bayesian variable selection
is using auxiliary indicator variables (1 indicating presence and 0
indicating absence of the corresponding covariate in the model) to
obtain a `spike and slab' prior on the regression coefficients
\cite[see e.g.][]{mitc:beau:1988, geor:mccu:1993, yuan:lin:2005,
  ishw:rao:2005, lian:paul:moli:clyd:berg:2008, john:ross:2012,
  rock:geor:2014, nari:he:2014, cast:schm:vand:2015, roy:tan:fleg:2018, shin:bhat:john:2018,
  li:dutt:roy:2023, barb:berg:2021}. Here the `spike' corresponds to the probability
mass concentrated at zero or around zero for the variables vulnerable
to deletion and the `slab' specifies prior uncertainty for
coefficients of other variables. We will discuss one such model in
details in Section~\ref{sec:hier}. Analysis using such models
determines (selects) the most promising variables by summarizing the
posterior density of the indicator variables and/or the regression
coefficients.

In the ultra-high dimensional settings where the number of variables
($p$) is much larger than the sample size ($n$), previously mentioned
variable selection methods may not work and the computational cost for
large-scale optimization or Markov chain Monte Carlo (MCMC)
exploration (in the Bayesian methods) becomes too high to afford. This
is why, in practice, a computationally inexpensive screening is
performed before conducting a refined model selection
analysis. Motivated by these, \cite{fan:lv:2008} proposed the sure
independence screening (SIS) method where marginal Pearson
correlations between the response and the variables are used to screen
out unimportant variables, and thus rapidly reduce the dimension to a
manageable size. The SIS method has been extended to generalized
linear models \citep{fan:song:2010} and additive models
\citep{fan:feng:song:2011} among others. Various other correlation
measures as for example general correlation \citep{hall:mill:2009},
distance correlation \citep{li:zhon:zhu:2012}, rank correlation
\citep{li:peng:zhan:zhu:2012}, tilted correlation
\citep{cho:fryz:2012, lin:pang:2014} and quantile partial correlation
\citep{ma:li:tsai:2017} have also been proposed to rank and screen
variables. \cite{chan:tang:wu:2013} discussed a marginal likelihood
ratio test, \cite{mai:zou:2013} used Kolmogorov-Smirnov statistic,
\cite{xu:chen:2014} suggested maximum likelihood estimate to remove
unimportant variables, respectively. \cite{he:wang:hong:2013}
discussed a nonparametric screening method, \cite{zhou:zhu:xu:li:2019}
used a divergence based screening method and \cite{mukh:duns:2020}
proposed the randomized independence screening. \cite{wang:2009}
studied the popular forward regression (FR) method \cite[see
also][]{hao:zhan:2014} and \cite{wang:leng:2016} proposed the high
dimensional ordinary least squares projection (HOLP) for screening
variables in ultra-high dimensional settings. However, currently, there is no 
available screening method based on a
Bayesian model that allows the incorporation of prior information on the
variables. On the other hand, in biological GWAS studies, such as the one 
considered in the paper, scientists often have prior knowledge on whether a 
trait is complex or simple. Scientists believe that complex traits result from 
variations within multiple genes while simple traits result from variations 
within a few genes with strong influence \cite{kale:gill:2020}. Such knowledge 
are used to elicit the prior on model size and effect size. In addition, prior 
knowledge on linkage disequilibrium can help choosing the shrinkage parameter. 
In this paper we develop a Bayesian screening method allowing the use of such 
prior information.

In order to develop a Bayesian screening method, we consider a
hierarchical model with zero inflated mixture priors which are special
cases of the spike and slab priors. As mentioned earlier, variants of
these hierarchical models have previously been used for variable
selection and MCMC algorithms are generally used to approximate
posterior probabilities. 
On the other hand, the sequential screening method proposed here called
{\it B}ayesian {\it it}erative {\it s}creening (BITS) does not involve any MCMC sampling.
Under the hierarchical model considered here, the marginal posterior probability mass function (pmf) of the latent indicator vector is analytically available up to a normalizing constant. BITS uses this pmf to iteratively include variables that have maximum posterior inclusion probability conditional on the already selected variables. The computation of the posterior probabilities is done by a one-step delayed Cholesky update. \cite{wang:leng:2016} mention that there are two important aspects of a successful screening method: computational efficiency and screening consistency property under flexible conditions. Even though BITS allows incorporation of prior information, it is computationally competitive with the frequentist screening
methods like HOLP. The hierarchical model on which BITS is based assumes Gaussian errors. But we show that BITS has screening consistency even under the more general $q$-exponential tail condition on the errors. Also, we do not assume that the marginal correlations for the important variables are bounded away from zero---an assumption that is often violated in practice but used for showing screening consistency of SIS. Finally, we introduce the notion of posterior screening consistency and discuss its usefulness in Bayesian high dimensional data analysis.

BITS although is similar in spirit to the popular, classical variable screening method, namely forward regression (FR) \citep{wang:2009} there are important differences between the two. By introducing the notion of ridge partial correlations, we show that unlike FR, BITS takes into account ridge partial variances to include potential variables. Also, by varying the ridge penalty, BITS can include groups of correlated important variables whereas FR selects only a candidate from each group which is not desirable for a screening method. For deciding the screened model size \cite{wang:2009} considers the extended BIC (EBIC) criterion developed by \citet{chen:chen:2008}. In addition to the use of EBIC and the liberal choice of having a model as large as the sample size, we construct a new stopping rule based on the posterior probability (PP). Furthermore, we prove that the PP stopping rule is screening consistent again under the general $q$-exponential tail condition on the errors. Through examples we demonstrate how the PP criterion can lead to informative screening by specifying suitable prior hyperparameter values.

The rest of the paper is organized as follows. We describe the
hierarchical model and BITS in Section~\ref{sec:scree}. We establish
screening consistency properties of BITS in Section
\ref{sec:consistency} as well as discuss the notion of posterior
screening consistency.  In Section~\ref{sec:comfw} we emphasize the
contrast between the proposed method and FR. In Section
\ref{sec:stopping} we describe different possible stopping rules for
BITS and establish screening consistency of the proposed PP stopping
rule. Section \ref{sec:otherprior} contains derivations of BITS for
some other priors. Section \ref{sec:sca} lays out the fast statistical
computation algorithm for BITS. Section \ref{sec:exam} contains
results from extensive simulation examples. In particular, these
examples are used to study the posterior mass coverage of BITS and to
compare BITS with other frequentist screening methods.  A real data
set from a genomewide association studies with more than half a
million markers is analyzed in Section \ref{sec:data}. Some concluding
remarks are given in Section~\ref{sec:disc}.  A supplement document
containing the proofs of the theoretical results and some further
simulation results is provided with sections referenced here with the
prefix `S'.  The methodology proposed here is implemented as a
function named \texttt{bits} in an accompanying \textsf{R} package
\texttt{bravo} \citep{bravo:2021}.

\section{A Bayesian iterative screening method}
\label{sec:scree}


\subsection{A hierarchical Gaussian regression model}
\label{sec:hier}


Let the vector $y=(y_1,\ldots,y_n)$ denote the $n\times 1$ vector of responses and the $n\times p$ matrix $X = (X_1,\ldots,X_p)$ denote the matrix of covariate values with vector of regression coefficients $\beta = (\beta_1,\dots,\beta_p).$ The Bayesian variable selection model we consider here assumes latent indicator vector $\gamma = (\gamma_1, \ldots, \gamma_p)^{\top} \in \{ 0, 1 \}^p,$  such that $X_j$ is included in the linear regression model if and only if $\gamma_j = 1$. 
Let $X_\gamma$ be the $n \times |\gamma|$ sub-matrix of $X$ that consists of columns of $X$ corresponding to model $\gamma,$ $\beta_\gamma$ be the vector that contains the regression coefficients for model $\gamma,$ and $|\gamma| = \sum_{i=1}^p \gamma_i$ be the model size. Without loss of generality, each column of $X$ is assumed standardized.  
We then assume the Bayesian Gaussian hierarchical model
\begin{subequations}
  \label{eq:vsblm}
  \begin{alignat}{2}
    \label{eq:vsblm-a}
y|\beta_0,\beta,\gamma,\sigma^2                  & \sim \sN_n(1_n \beta_0 + X_{\gamma} \beta_{\gamma},\sigma^2I_n),               \\
\label{eq:vsblm-b}
 (\sigma^2, \beta_0) & \sim f(\beta_0, \sigma^2) \propto 1 / \sigma^2,                                       \\
    \label{eq:vsblm-c}
  \beta_j | \gamma,\sigma^2    & \ind {\cal N} \bigl( 0, \frac{\gamma_j}{\lambda} \sigma^2  \bigr) \;\;\text{for $j=1,\ldots,p$}, \\
    \label{eq:vsblm-d}
  \gamma | w& \sim f(\gamma|w) = w^{|{\gamma}|} (1 - w)^{p - |{\gamma}|}.
  \end{alignat}
\end{subequations}
In this hierarchical setup a popular non-informative prior is set for $(\beta_0,\sigma^2)$ in \eqref{eq:vsblm-b} and a conjugate independent normal prior is used on $\beta$ given $\gamma$ in \eqref{eq:vsblm-c} with $\lambda > 0$ controlling the precision of the prior independently from the scales of measurements. Note that under this prior, if a covariate is not included in the model, the prior on the corresponding regression coefficient degenerates at zero. In \eqref{eq:vsblm-d} an independent Bernoulli prior is set for $\gamma$, where $w \in (0, 1)$ reflects the prior inclusion
probability of each predictor. We assume $\lambda$ and $w$ are known non-random functions of $n$ and $p$.

Our Bayesian screening method hinges on the fact that given $\bm\gamma,$ it is possible to integrate out other variables analytically. Indeed, integrating out $(\beta_\gamma, \beta_0,\sigma^2)$ 
we derive the following marginal distribution of $y$ given $\gamma$, 
\begin{eqnarray}
 f(y |\gamma)  &=&  \int_{{\Real}_+} \int_\Real \int_{\Real^{|\gamma|}} f(y|\gamma,  \sigma^2, \beta_0, \beta_{\gamma}) f(\beta_{\gamma}|\gamma,  \sigma^2, \beta_0) f(\sigma^2, \beta_0) d \beta_{\gamma} d\beta_0 d {\sigma^2} \nonumber\\
   \label{eq:vsb-a}
&  = &c_n \,\lambda^{|\gamma|/2} \big|X_{\gamma}^{\top} X_{\gamma}+ \lambda I  \big|^{-1/2}
 \bigl[  \tilde{y}^{\top} \tilde{y}  -\widetilde{\beta}_\gamma^{\top} (X_{\gamma}^{\top} X_{\gamma}+ \lambda I) \widetilde{\beta}_\gamma \bigr]^{-(n-1)/2},
\end{eqnarray}
where, $c_n= \Gamma((n-1)/2)/\pi^{(n-1)/2}$ is a constant depending only on the sample size $n$,
$\tilde{y} = y - \bar{y} 1_n$, and
$\widetilde{\beta}_\gamma= (X_{\gamma}^{\top} X_{\gamma}+ \lambda I
\big)^{-1} X_{\gamma}^{\top} \tilde{y}$. 
The hierarchical Gaussian regression model \eqref{eq:vsblm} and its variants have been used extensively and exclusively for variable selection. In particular, several works have established strong model selection consistency results \citep{nari:he:2014} under the ultra-high dimensional setup that is considered here. In practice, however, these methods are mostly used after reducing the number of covariates using frequentist screening methods that are not driven by the same Bayesian hierarchical model. In the next section we describe a novel Bayesian screening method based on the hierarchical model \eqref{eq:vsblm}.


\subsection{The BITS algorithm}
\label{sec:bis}


We now describe our proposed screening method. This method uses the
posterior pmf of $\gamma$, $f(\gamma | y)$ which is available up
to a normalizing constant. Indeed, if $w$ is assumed
fixed, then
\begin{equation}
  \label{eq:logpostgam}
\begin{split}
  &\log f(\gamma | y) = \mbox{const}  + \frac{|\gamma|}{2} \log \lambda - \frac{1}{2} \log \big|X_{\gamma}^{\top} X_{\gamma} + \lambda I  \big|\\
& \quad - \frac{n-1}{2} \log \bigl[  \tilde{y}^{\top} \tilde{y}  - \tilde{y}^{\top}X_{\gamma} (X_{\gamma}^{\top} X_{\gamma}+ \lambda I
\big)^{-1} X_{\gamma}^{\top} \tilde{y} \bigr]+ |\gamma| \log\frac{w}{1-w}.
\end{split}
\end{equation}
Let $e_i$ be the $i$th $p$ dimensional canonical basis vector, that is, the $i$th
element of $e_i$ is one, and all other elements are zero. In the first step,
we select the variable $i_1$ such that 
$i_1 = \argmax_{j \in \{1, \dots, p\}} \log P(\gamma = e_j |y).$   
Thus we select the unit vector $e_j$ with highest posterior probability. In the next
step, we select the variable $i_2$ with
$i_2 = \argmax_{j \neq i_1} \log P(\gamma = e_{i_1} + e_j |y).$ 
Note that for $j \neq i_1$,
\begin{equation}\label{eq:equivalence}
 P(\gamma_j =1 | \gamma_{i_1} = 1, \gamma_k =0, k \neq i_1; y) =\Bigg(1+\frac{P(\gamma = e_{i_1} |y)}{P(\gamma = e_{i_1} + e_j |y)}\Bigg)^{-1}.
\end{equation}
So, $P(\gamma_j =1 | \gamma_{i_1} = 1, \gamma_k =0, k \neq i_1; y)$ and $P(\gamma = e_{i_1} + e_j |y)$ are maximized at the same $j \neq i_1$. Thus, in the second step, we choose the variable which has maximum
posterior inclusion probability given that $i_1$ (the variable
selected in the first step) is included in the model. An efficient
computational method for calculating $P(\gamma = e_{i_1} + e_j |y)$
using a fast Cholesky update is given in Section \ref{sec:sca}. One problem
with marginal correlation based screening methods like SIS
\citep{fan:lv:2008} is that unimportant variables that are correlated
with important variables may get selected. This is not likely to
happen in our proposed screening procedure as the only variables that
have high (conditional) posterior inclusion probabilities {\it after}
taking into account the selected variables survive the
screening. 
Let
$\gamma^{(m)} = \sum_{k=1}^m e_{i_j}$ be the $\gamma$ vector at the
$m$th step.
Below we describe the $(m+1)$st iteration of the BITS.  \baro \vspace*{2mm}
\noindent {\rm Iteration $m+1$ of the screening algorithm:}

Given $\gamma^{(m)}$, let
\begin{equation}\label{eqn:bits-general}
 i_{m+1} = \argmax_{j \notin \gamma^{(m)}} \log P(\gamma = \gamma^{(m)} + e_j |y).
\end{equation}
Set $\gamma^{(m+1)}= \gamma^{(m)} + e_{i_{m+1}}$.
\barba \bigskip
Using the same argument as \eqref{eq:equivalence}, the $i_{m+1}$st variable has the highest posterior probability of being included given that $i_1,i_2,\ldots,i_m$ have already been included in the model. BITS although is not guaranteed to produce the posterior mode, or any other standard summary measures of the posterior distribution $f(\gamma | y)$, in the next section we show that it enjoys screening consistency. 

\begin{remark}
\label{rem:sis}
Since, $P(\gamma = e_j |y) \propto [1  - r_j^2/(1+\lambda/n)]^{-(n-1)/2},$
with $r_j = X_j^{\top}\tilde{y}$, the first step of the BITS algorithm 
selects the variable with largest marginal correlation as in \pcite{fan:lv:2008} SIS algorithm.
\end{remark}
\begin{remark}
 Under the Bernoulli prior \eqref{eq:vsblm-d} on $\gamma$ the selection path $i_1,i_2,i_3,\ldots$ does not depend on the hyperparameter $w$ because the prior is only a function of $|\gamma|.$ However, as we shall see later in this section, $w$ has an effect on some stopping rules and the screening consistency.
\end{remark}

\subsection{Screening consistency of BITS}\label{sec:consistency}


Ideally, as the sample size increases we would like all the important variables to be included in $\gamma^{(m)}$ after a reasonable number of $m (< n)$ steps. The notion of frequentist screening consistency \citep{fan:lv:2008,wang:2009} states that if $E(y) = X_t\beta_t,$ for some subset $t\subset \{1,\ldots,p\}$ then $P(t \in \gamma^{(m)} \textrm{ for some } m \le n)$ should converge to $1$ as $n\to\infty,$ under some regularity conditions. In order to state the assumptions and the results more rigorously, we use the following notations. Abusing notation, we interchangeably use a model $\gamma$ either as a $p$-dimensional
binary vector or as a set of indices of non-zero entries of the binary
vector. For models $\gamma$ and $s,$ $\gamma^c$ denotes the complement
of the model $\gamma$. For $1\le i \le p,$ we say $i\in\gamma$ if $\gamma_i = 1$ and $i\notin\gamma$ if $\gamma_i=0$. For two positive
real sequences $\{a_n\}$ and $\{b_n\}$, $a_n \sim b_n$ means
${a_n}/{b_n}\rightarrow c$ for some constant $c>0$; $a_n \succeq b_n$
(or $b_n \preceq a_n$) means $b_n = O(a_n)$; $a_n \succ b_n$ (or
$b_n \prec a_n$) means $b_n = o(a_n)$. Again, abusing notations, for two real
numbers $a$ and $b$, $a \vee b$ and $a \wedge b$ denote max$(a, b)$ and min$(a, b),$ respectively. Finally, let $\beta_+ = \min_{i\in t}|\beta_i|.$ We first consider the orthogonal design case in Section~\ref{sec:ortho} and then more general design matrices and misspecified errors in Section~\ref{sec:genscr}.


\subsubsection{Orthogonal design with Gaussian errors}\label{sec:ortho}


The following theorem shows that under a Gaussianity assumption and some mild conditions on the effects sizes, BITS include all and only the important variables in the first $|t|$ steps.
\begin{theorem}\label{thm:orthogonal}
 Suppose $X^\top X = nI_p,$ $y = \beta_01_n + X_t\beta_t + \sigma \epsilon$ where $\epsilon\sim \sN(0,I_n),$ $|t| = O(n^a),$ for some $a < 1,$ and $n\beta_+^2 \succ \log n.$ Then for any $\lambda > 0,$ $P(\gamma^{(|t|)} = t) \to 1$ as $n\to \infty.$
\end{theorem}

Such a strong conclusion holds for the orthogonal design because the variables are uncorrelated and hence the marginal correlations are asymptotically ordered by the magnitudes of the regression coefficients \citep{fan:lv:2008}. Furthermore, the orthogonality restricts $p$ to be at most $n$. Thus, it is unrealistic to expect the same conclusion to hold in general situations. In particular, note that the first variable included is the one with the highest absolute marginal correlation with the response (Remark~\ref{rem:sis}). There are ample examples of realistic designs \citep{fan:lv:2008,wang:dutt:roy:2020} where an unimportant variable has the highest absolute marginal correlation with the response.


\subsubsection{Screening consistency in more general cases}\label{sec:genscr}


Although BITS is developed under the Gaussianity assumption on $y$ for computational tractability, we would like to establish screening consistency even under more general tail conditions. As we shall see, the tail behavior of the error distribution plays a crucial role in proving screening consistency. To that end, we consider the family of distributions with $q$-exponential tail condition \citep{wang:leng:2016} given below.
\begin{definition}[$q$-exponential tail condition]
A zero-mean distribution $F$ is said to have $q$-exponential tail, if there exists a function $q:[0,\infty)\to \mathbb{R}$ such that for any $N\geq 1,$ $\eta_1,\ldots,\eta_N \stackrel{iid}{\sim} F$, $\ell\in{\cal R}^N$ with $\|\ell\| = 1$, and $\zeta > 0$ we have $P\left(\left|\ell^\top \eta\right| > \zeta\right)\leq\exp\{1-q(\zeta)\}$ where $\eta = (\eta_1,\ldots,\eta_N).$
\end{definition}
This tail condition is assumed by \cite{wang:leng:2016} in establishing screening consistency of their HOLP screening method. In particular, as shown in \cite{vers:2012}, for standard normal distribution, $q(\zeta) = \zeta^2/2,$ when $F$ is sub-Gaussian $q(\zeta) = c_F\zeta^2$ for some constant $c_F > 0$ depending on $F,$ and when $F$ is sub-exponential distribution  $q(\zeta) = c'_F\zeta$ for some constant $c'_F > 0$ depending on $F.$ Finally, when only first $2k$ moments of $F$ are finite, $q(\zeta) - 2k\log(\zeta) = O(1).$

We assume the following set of conditions:
\begin{enumerate}
\item[C1] $y=\beta_01_n + X_t\beta_{t} + \sigma\epsilon$ where $t$ is the true model, $\epsilon=(\epsilon_1,\dots, \epsilon_n)$, $\epsilon_i \stackrel{iid}{\sim} F_0$ which has $q$-exponential tail with unit variance.
\item[C2] $|\log\lambda| = O(\log n)$. 
\item[C3] There exist $K_n$ and $0 < \delta < 1/\sigma$ such that $(K_n+1)|t|\leq n$,
\[\dfrac{|t|\|\beta_t\|^2\log n }{\tau_{+}^2\beta_{+}^4} \prec K_n \preceq n\log n \min\left\{\frac{\tau_+}{\lambda}, \frac{1}{|\log (1/w-1)|} \right\} \]
\[ q(s_n\delta) - K_n|t|\log p - \log|t|\to\infty\]
 where $s_n = \sqrt{n}\tau_+\beta_+^2/(\|\beta_t\|\sqrt{|t|})$ and $\tau_+$ is the smallest nonzero eigenvalue of $X_\gamma^{\top}X_\gamma/n$ with $|\gamma|\leq (K_n+1)|t|.$

\end{enumerate}

Although our assumption on $p$ is related only to the tail behavior of the error 
distribution, it is evident that under sub-Gaussian or sub-exponential tailed 
$F_0$, BITS is screening consistent in the ultra-high dimensional setting. For 
example, suppose $\|\beta_t\| = O(1),$ $\tau_+ \sim n^{-h_1}$, $\beta_+ \sim 
n^{-h_2},$ $\log p \sim n^{h_3},$ $|t|\sim n^{h_4}, w = n^{-h_5}, $ for some 
$h_i >0, i=1,\dots,5,$ and  $K_n=n^{2h_1 +4h_2 +h_4}(\log n)^2.$ Note that, 
these values of $\beta_+$ and $|t|$ are as in \cite{wang:leng:2016}. Also, 
suppose $\lambda = n^{h_6}$ for some constant $h_6$. When $F_0$ is sub-Gaussian, 
simple calculations show that a sufficient condition for (C1)--(C3) is that 
$4h_1+8h_2+h_3+3h_4 + h_6<1$ if $h_1+h_6 >0$ and $4h_1+8h_2+h_3+3h_4 <1$ if 
$h_1+h_6 \le 0.$ If $F_0$ is subexponential, a sufficient condition is 
$6h_1+12h_2+2h_3+5h_4 + h_6<1$ if $h_1+h_6 >0$ and $6h_1+12h_2+2h_3+5h_4 <1$ if 
$h_1+h_6 \le 0,$ which is slightly more stringent than the sub-Gaussian case. 
Furthermore, under any of these sufficient conditions, $K_n |t| \prec \lfloor 
n/log n \rfloor$ which is typically the screened model size of SIS 
\citep{fan:samw:wu:2009}.
Note that, C2 assumes quite a weak condition on the prior shrinkage parameter
$\lambda$, allowing us to run BITS for different choices of $\lambda$,
and uniting these results to produce the screened model. Condition C3 lays out the explicit scaling law of $w$ against $|t|$ for screening consistency of BITS and the PP stopping rule.
We now present the screening consistency result. 
\begin{theorem}
\label{thm:consistency}
 Under conditions (C1)--(C3), there exists $c > 0$ such that for all sufficiently large $n,$ 
 \[P(\gamma^{(K_n|t|)} \supseteq t) \ge 1 - \exp(1 - q(s_n\delta) + K_n|t|\log p + \log|t|) - P(\|\tilde{y}\|^2 >n u_n),\]
 where $u_n = \tau_+^2\beta_+^4K_n/(c|t|\|\beta_t\|^2\log n).$
\end{theorem}

Note that, the limit of $(\|\tilde{y}\|^2 -\beta_t^\top X_t^\top X_t \beta_t)/n$ is bounded above by $ \sigma^2$ almost everywhere. Thus, if $\beta_t^\top X_t^\top X_t \beta_t/n \prec u_n$, it is easy to see from C3 that $P\left(\gamma^{(K_n|t|)}\supseteq t\right) \to 1.$ That is, with overwhelmingly large probability, the true model is included in at most $K_n|t|$ many steps. Note that, the condition $\beta_t^\top X_t^\top X_t \beta_t/n \prec u_n$ is weaker than the condtion var ($y) = O(1)$ assumed in \cite{wang:leng:2016}. However, when higher order moments of $\epsilon_1$ exist, further lower bounds to $P(\gamma^{(K_n|t|)} \supseteq t)$ can be obtained as described in the following corollary.

\begin{corollary}
  \label{cor:consistency}
  Suppose conditions (C1)--(C3) holds. If further, $\beta_t^\top X_t^\top X_t \beta_t/n \prec u_n \wedge n$ and $E(\epsilon_1^6) < \infty$ then with $\kappa =$ Var$(\epsilon_1^2)$, $v_n = u_n -\beta_t^\top X_t^\top X_t \beta_t/n -1$ and a constant $c_1 > 0,$ for all sufficiently large $n,$ $$P(\|\tilde{y}\|^2 >n u_n) < \frac{\sigma^2 \sqrt{\kappa}}{\sqrt{2n\pi} v_n}\exp\left(-\dfrac{nv_n^2}{2\sigma^4\kappa}\right) + \dfrac{c_1}{\sqrt{n}} + \frac{4\sigma^4 \beta_{t}^{\top}X_{t}^{\top}X_{t}\beta_{t}}{n^2}.$$
\end{corollary}


\subsubsection{Posterior screening consistency}
\label{sec:postscr}


The screening consistency considered in Sections~\ref{sec:ortho} and
\ref{sec:genscr} are in the frequentist sense and therefore are not
guaranteed to be fidelitous to the posterior inference.  In this
section we discuss the concept of posterior screening consistency
\citep[see also][Theorem 3]{song:lian:2015}. We start with the
following definition.
\begin{definition}
\label{def:bscreen}
A sequence of models $\{\gamma^{[n]}\},$ with $|\gamma^{[n]}| \leq n$ is said to be posterior screening consistent if 
\begin{equation}
    \label{eq:bayeconsistent}
P\left(\gamma \in {\cal P}_t(\gamma^{[n]}) \big|y\right) \equiv \sum_{t\subseteq\gamma \subseteq \gamma^{[n]}} f(\gamma|y) \to 1
\end{equation}
in probability as $n\rightarrow\infty,$ where ${\cal P}_t(\gamma^{[n]})$ is the set of all sub-models of $\gamma^{[n]}$ containing $t.$
\end{definition}
In other words, with probability tending to 1, the posterior mass of $\gamma$ is entirely supported on $2^{(|\gamma^{[n]}|-|t|)}$ models which are sub-models of $\gamma^{[n]}.$ Considering that there are originally $2^p$ possible models, this can be a great reduction in the search space for models with high posterior probabilities.
Typically Bayesian variable selection algorithms search for the
posterior mode and other high-posterior probability models. Many competitive algorithms are available to
search for the best model including, but surely not limited to, the
stochastic shotgun algorithm \citep{hans:dobr:2007}, simplified
stochastic shotgun algorithm \citep{shin:bhat:john:2018}, shotgun with embedded screening \citep{li:dutt:roy:2023}, Gibbs
sampling \citep{nari:he:2014}, Metropolis-Hastings algorithm
\citep{zhou:guan:2019} and mixed integer optimization \citep{bert:king:2016}. Since the size ($2^p$) of the model space
grows exponentially with the number of variables, due to computational
cost and convergence issues of these iterative algorithms, when dealing with high dimensional data sets,
generally a screening step is performed before applying a Bayesian
variable selection algorithm. For example, \cite{nari:he:2014}, in
their real data example, use SIS to reduce the number of variables
 from 22,575 to 400 before applying their variable selection
algorithm. One important aspect of posterior screening consistency is
that the best model (in terms of posterior probability) for variable
selection can be searched among a much smaller number of models
instead of among the humongous number ($2^p$) of all possible
models. Thus a posterior screening consistent model can serve as an excellent starting point for implementing further Bayesian variable selection methods.

In the above we have described important practical consequences of
using posterior screening consistent algorithms. We now discuss
conditions guaranteeing such consistency. In the context of ultra-high
dimensional Bayesian variable selection, under different hierarchical model setups, recently several articles
have established strong (posterior) selection consistency, that is,
$f(t|y)\to1$ in probability as $n\to\infty$ \cite[see
e.g.][]{nari:he:2014, yang:wain:jord:2016, shin:bhat:john:2018,
li:dutt:roy:2023}. Thus under strong selection consistency, posterior probability of the true model goes to one as $n \rightarrow \infty$. Now, for given $\delta > 0$, denoting the events
  $\{t \subseteq \gamma^{[n]}\}$ and $\{f(t|y) > 1-\delta\}$ by
  $A_n$ and $B_{n,\delta}$ respectively, we have
    $P(\sum_{t\subseteq\gamma \subseteq \gamma^{[n]}} f(\gamma|y) > 1 - \delta) \ge P(A_n \cap B_{n,\delta}) \ge P(A_n)+ P(B_{n,\delta}) -1.$
 Thus, if a sequence of models $\{\gamma^{[n]}\}$ is screening consistent
  and strong selection consistency holds, that is, if $P(A_n)\rightarrow 1$ and $P(B_{n,\delta}) \rightarrow 1$
  as $n \rightarrow \infty$ then $\{\gamma^{[n]}\}$ is posterior screening consistent. \cite{yang:wain:jord:2016} and \cite{li:dutt:roy:2023} derive conditions for strong selection consistency for the model \eqref{eq:vsblm}. So, by Theorem~\ref{thm:consistency} we have posterior screening consistency results for BITS.

  \subsection{Comparison with frequentist screening methods}
  In this Section we compare BITS with two frequentist screening
  methods, namely \pcite{wang:2009} FR and
  \pcite{wang:leng:2016} HOLP.
  
\subsubsection{Contrast with the forward regression method}
\label{sec:comfw}


BITS although is similar in spirit to the forward selection, the
step-wise regression method, there are significant differences between
BITS and the FR method of \cite{wang:2009}. Firstly, under the conditions of \citet{wang:2009}, BITS is screening consistent:
\begin{lemma}
\label{lem:Kexist}
Under the conditions of \cite{wang:2009}, that is, with $\beta_+\geq \nu_{\beta}n^{-\xi_{min}}, \log p\leq \nu n^{\xi}, |t|\leq \nu n^{\xi_0}, \|\beta_t\|\leq C_{\beta}$ for some finite constant $C_{\beta}$, $\tau_+\geq \tau_{{min}}$ for some finite constant $\tau_{{min}}$, and $\xi + 6\xi_0 + 12\xi_{min}<1$, $\epsilon_i \stackrel{iid}{\sim}{\cal N}(0,1)$, then C3 holds with $K_n = n^{\xi_0+4\xi_{min}} (\log n)^2$.
\end{lemma}

A proof of Lemma~\ref{lem:Kexist} is given in Section \ref{sec:Kexist}
of the supplement. We indeed prove that the lemma holds under a weaker
condition of $\xi + 3\xi_0 + 8\xi_{min}<1.$ In order to show the
contrasts between BITS and FR, we introduce the notion of ridge
partial correlations.
\begin{definition}\label{def.ridgepartials}
For any $\gamma$ and $i \notin \gamma$, the ridge partial correlation between $y$ and $X_i$ given $X_\gamma$ with ridge penalty $\lambda$ is given by
\[R_{i\cdot\gamma,\lambda} \equiv R_{iy\cdot\gamma,\lambda} = -
  v_{iy\cdot\gamma,\lambda}\big/\left\{v_{i\cdot\gamma,\lambda}\times
    v_{y\cdot\gamma,\lambda}\right\}^{1/2}\] where
$v_{i\cdot\gamma,\lambda} = n^{-1} X_i^\top X_i +n^{-1}\lambda- n^{-2}
X_i^\top X_\gamma\left(n^{-1}X_\gamma^\top X_\gamma + \lambda/n
  I\right)^{-1}X_\gamma^\top X_i$ is the ridge (sample) partial
variance of $X_i$ given $X_\gamma$,
$v_{iy\cdot\gamma,\lambda} = n^{-1}\tilde{y}^\top X_i -
n^{-2}\tilde{y}^\top X_\gamma \left(n^{-1}X_{\gamma}^\top X_{\gamma} +
  \lambda/n I\right)^{-1}X_\gamma^\top X_i,$ and
\[v_{y\cdot\gamma,\lambda} = n^{-1}\tilde{y}^\top\tilde{y} -
n^{-2}\tilde{y}^\top X_\gamma\left(n^{-1}X_\gamma^\top X_\gamma +
  \lambda/n I\right)^{-1}X_\gamma^\top\tilde{y}.\]
\end{definition}
\begin{remark}
 When $|\gamma| < n$ and $\lambda = 0,$ $R_{i\cdot\gamma,0}$
is exactly the (sample) partial correlation between $y$ and
$ X_i$ after eliminating the effects of
$ X_j, j \in \gamma$, and the ridge partial sample
variances $v_{i\cdot\gamma,0}$'s are exactly the (sample) partial
variances of $ X_i$'s given
$ X_j, j \in \gamma$.
\end{remark}

In fact, BITS can be reformulated using these ridge sample partial
correlations and ridge sample partial variances. To that end, we first
show how the log-marginal posterior probability increments depend on the ridge
partial correlations and partial variances.
\begin{lemma}\label{lemma:fw1}
 For the model \eqref{eq:vsblm}, for any $\gamma,$ and $i \notin \gamma,$ 
 \[\frac{f(\gamma + e_i|y)}{f(\gamma |y)} = \frac{w(n\lambda)^{1/2}}{(1-w)v_{i\cdot\gamma,\lambda}^{1/2}(1 - R^2_{i\cdot\gamma,\lambda})^{(n-1)/2}}.
   \]
\end{lemma}

Consequently, under the
independent prior, having chosen model $\gamma^{(m)},$ the BITS method
chooses the candidate index $i_{m+1}$ that maximizes
$-\shalf\log v_{i\cdot\gamma^{(m)},\lambda} - \shalf(n-1)\log \{ 1 -
R^2_{i\cdot\gamma^{(m)},\lambda}\}$ over $i \notin \gamma.$ This
is clearly different from the forward screening method of
\cite{wang:2009} where it only maximizes the absolute partial
correlations $|R_{i\cdot\gamma,0}|,$ $(i\notin \gamma).$ In
particular, BITS also takes into consideration the (sample) partial
variances of each $X_i$ given the already included variables. Thus,
between two candidates which have the same partial correlations with
the response given already included variables, the one with smaller
conditional variance is preferred. However, if the partial
correlations with the response are different then because of the
presence of the the multiplier $(n-1),$ the effect of the conditional
variance is practically insignificant. Furthermore, by shrinking the effects using the ridge penalty, BITS can include groups of important variables that are highly correlated among themselves in contrast to FR which would only select a candidate variable from the group. Finally, note that
during screening it could be useful to be liberal and include more
than $n$ variables. This however, is not possible by FR because all
models of size bigger than $n-1$ have \emph{zero} residual sum of
squares. BITS, on the other hand, allows to have a screened model of
size bigger than $n.$ In different simulation examples in
section~\ref{sec:exam}, we demonstrate that BITS performs much better than FR.

\subsubsection{Contrast with the HOLP}
\label{sec:comholp}
For fixed, $\gamma, \sigma^2$, the mode of the posterior density of
$\beta_\gamma$ under the prior \eqref{eq:vsblm-c} is
$\widetilde{\beta}_\gamma$ mentioned in Section~\ref{sec:hier}. The
ridge regression estimate of $\beta$ corresponding to the ridge
penalty $\lambda$ is
$\hat{\beta}(\lambda) = (X^{\top} X+ \lambda I \big)^{-1} X^{\top}
\tilde{y}$. Thus, for given $\gamma$, $\widetilde{\beta}_\gamma$ is
the same as the ridge regression estimate. \cite{wang:leng:2016}
showed that the HOLP estimator can be obtained as
$\lim_{\lambda \rightarrow 0} \hat{\beta}(\lambda)$. The posterior
probability $f(\gamma |y)$ given in \eqref{eq:logpostgam} that steers
the BITS algorithm involves $\widetilde{\beta}_\gamma$. While BITS
sequentially selects the variables based on \eqref{eq:logpostgam}, in
their numerical examples \cite{wang:leng:2016} use either the
variables with $n$ largest (absolute) HOLP estimates or the EBIC
method \citep{chen:chen:2008} to implement the HOLP method. In the
next Section, we provide a posterior probability based consistent
stopping rule for BITS.

\subsection{Stopping criteria}\label{sec:stopping}


Note that BITS provide a sequence of predictor indices $i_1,i_2,\ldots.$ A practical question is when to stop the algorithm. Theorem \ref{thm:consistency} suggests that the first $K_n|t|$ indices contain the true model with overwhelming probability. However, there is nothing to stop us from being liberal and include the first $n$ indices $i_1,i_2,\ldots,i_n.$ 

The aforementioned rule is not affected by the prior inclusion
probability $w.$ We now propose a \textit{new} stopping rule, called
the posterior probability (PP) criterion that depends on $w$.  As we
expect the important variables to be included early, the BITS
algorithm is expected to provide a sequence of nested models
$\{\gamma^{(i)}\}$ with increasing posterior probabilities until all
the important variables are included. Thus we may stop BITS when the
first drop occurs in these posterior probabilities, that is, we stop
at iteration
\[ {\cal T} = \arg\min_{m<K_n|t|}\{f(\gamma^{(m+1)}|y) < f(\gamma^{(m)}|y)\}. \]
Since $K_n|t| <n$, the PP criterion never selects more than $n$
variables. Also, the PP rule, just like the BITS method, is not searching
for the mode of the posterior pmf $f(\gamma|y)$. Note that BITS does not prune variables. On the other hand, finding a local mode would require
considering forward, backward, swap, and potentially other moves \citep{li:dutt:roy:2023}. Under the orthogonal design described in Section
\ref{sec:ortho}, we saw that with probability tending to one, BITS
include all and only the important variables in the first $|t|$
steps. In addition, we will now prove that the PP criterion stops
right after $|t|$ steps.
\begin{theorem}\label{thm.ortho_stop}
If the conditions for Theorem \ref{thm:orthogonal} hold and in addition, that $\|\beta_t\| = O(1)$ and that there exists $1/2 < c < 1$ such that $w = c'n^{-c}$ for some $c'>0$. Then
  $P(\mathcal{T} = |t|) \to 1$ as $n\to\infty.$
\end{theorem}

However, in general, as we have seen in Section \ref{sec:consistency}, BITS may include unimportant variables before it includes all the important variables. Thus it is unrealistic to require that the conclusions of the previous theorem holds in the more general case. However, the following theorem guarantees that $\gamma^{(\cal T)}$ is screening consistent.
\begin{theorem}
  \label{thm.gen_stop}
Under conditions (C1)--(C3), there exists a positive constant $c^*$ such that for all sufficiently large $n,$ 
$$P(\gamma^{(\cal T)} \supseteq t) \ge 1 - \exp(1 - q(s_n\delta) + K_n|t|\log p + \log|t|) - P(\|\tilde{y}\|^2 > nu_n^*),$$
 where $u_n^* = \tau_+^2\beta_+^4K_n/(c^*|t|\|\beta_t\|^2\log n).$
\end{theorem}

Recall that $\tilde{y}$ is the mean-centered $y$, that is,
$\tilde{y}= y- \bar{y}1_n$. Also, following the discussions right after
Theorem \ref{thm:consistency} and Corollary \ref{cor:consistency}, an upper bound to the last term
$P(\|\tilde{y}\|^2 > nu_n^*)$ can be obtained. This results in lower
bounds to $P(\gamma^{(\cal T)} \supseteq t)$ and in particular, establishing
$P(\gamma^{(\cal T)} \supseteq t) \rightarrow 1$.

Alternatively, \citet{wang:2009} and \citet{wang:leng:2016} also
promote the use of EBIC \citep{chen:chen:2008}. Under this stopping
rule, the screening is stopped at the smallest EBIC, that is, at
$\argmin_{1 \le m \le n-1} \textrm{EBIC}(m)$ where for $1\leq k < n$,
$\textrm{EBIC}(k) = \log\hat{\sigma}^2_{(k)} + k(\log n +2\log p)/n,$
and $n\hat{\sigma}^2_{(k)}$ is the ordinary least squares residual sum
of squares from regression of $y$ on the first $k$ screened
variables. Evidently, due to its ultra-high dimensional penalty, EBIC
is expected to be very conservative and yield small screened
models. Compared to the PP rule, the EBIC rule is computationally
expensive as it requires a model of size $n-1$.  On the other hand, a
similar variant of the PP criterion can be to choose the model
according to the largest drop in the posterior probability among the
first $K_n|t|$ steps, that is, the model size is
$\argmax_{1 \le m \le K_n|t|} (f(\gamma^{(m)}|y)
-f(\gamma^{(m+1)}|y)).$

\subsection{BITS for other priors}
\label{sec:otherprior}
A popular alternative to the independent normal prior in \eqref{eq:vsblm-c} is the Zellner's $g$-prior \citep{zell:1986} for
$\beta_\gamma$ indexed by a hyperparameter $g$ under the assumption that all $n\times n$ submatrices of $X$ are non-singular. That is, we also
consider the hierarchical model \eqref{eq:vsblm} where the prior in
\eqref{eq:vsblm-c} is replaced with
 $ \beta_{\gamma} | \gamma, \sigma^2 \sim {\cal N}_{|\gamma|} \bigl( 0, g\sigma^2(X_{\gamma}^{\top} X_{\gamma})^{-1}\bigr),$
 whence the marginal
density $\tilde{f}(y|\gamma)$ becomes
\begin{equation}
  \label{eq:zellmar}
       \tilde{f}(y|\gamma) = c_n \,(g+1)^{-|\gamma|/2} 
 \Bigl[  \tilde{y}^{\top} \tilde{y}  - \frac{g}{g+1}\tilde{y}^{\top} X_{\gamma}(X_{\gamma}^{\top} X_{\gamma})^{-1} X_{\gamma}^{\top} \tilde{y}\Bigr]^{-(n-1)/2}. \nonumber
\end{equation}
It is evident that the Zellner's $g$-prior provides the same screening path as FR because for fixed $|\gamma|$ the posterior pmf $\tilde{f}(\gamma|y)$ under Zellner's prior is a monotonic decreasing function of the regression sum of squares $\tilde{y}^{\top} X_{\gamma}(X_{\gamma}^{\top} X_{\gamma})^{-1} X_{\gamma}^{\top} \tilde{y}.$ Also, the Zellner's $g$-prior does not allow to have screened models of size more than $n.$

Similarly, BITS can easily accommodate beta-binomial prior distribution on $\gamma:$
$p(\gamma|a,b) = {\cal B}(|\gamma|+a,p-|\gamma|+b)/{\cal B}(a,b),$ where $\cal B$ is the beta function, and $a,b > 0.$ Since the beta-binomial prior also depend on $\gamma$ only via $|\gamma|,$ it does not have any effect on the screening path of BITS for a given $\lambda.$

Recently, \citet{koji:koma:2016} have proposed a class of discrete determinantal point process priors on the model space that discourages simultaneous selection of collinear predictors. The founding member of this class of priors is given by
$f_d(\gamma) = \left|d X^\top_\gamma X_\gamma\right|/\left|d X^\top X + I_p\right|,$
 where $d > 0$ controls the prior expectation of the model size. A value of $d > 1$ promotes larger models, while $d \leq 1$ promotes smaller models. Although \citet{koji:koma:2016} have studied the prior when $p<n,$ it can be also used when $p>n.$ In particular, $f_d(\gamma)$ puts zero prior probability on all models of size greater than $n.$ Notice that this prior is not a function of $|\gamma|$ and hence $d$ will have an effect on the BITS screening path.


\section{Fast statistical computation}
\label{sec:sca}


In this section we describe how BITS is implemented in practice. One major challenge in BITS is the computation of posterior probabilities of  models $\gamma^{(m)}+e_j$
for all $j\notin \gamma^{(m)}$ in the $(m+1)$st iteration. We show how these can be computed in minimal computational complexity using a one-step delayed Cholesky updates.

Before delving into the algorithm, let us define the notation $\odot$ and $\oslash$ as the element-wise multiplication and division between two vectors. Also when adding or subtracting a scalar to or from each entry of a vector we use the traditional `$+$' and `$-$' operators. The original covariates may have unbalanced scales and the scaled covariate matrix $X$ is often a dense matrix. Let $Z$ be the $n\times p$ matrix of the original covariates. We denote by ${r}$ the vector $X^\top \tilde{y}$ which can be simply computed as $D^{-1/2}Z^\top\tilde{y}$ without having to store $X.$ Also for greater numerical stability, we scale the vector $\tilde{y}$ so that $\|\tilde{y}\|^2= n,$ although the algorithm is described without this assumption. 

In the first iteration, $i_1 = \arg\max_i{r}_i$ and thus $\gamma^{(1)} = \{i_1\}.$ Let $R_1 \equiv b_1 = \sqrt{n+\lambda},$ denote the Cholesky factor of $X_{\gamma^{(1)}}^\top X_{\gamma^{(1)}} + \lambda I_1.$ Also let $v_{1} = {r}_{i_1}/b_1$ and let
\[\pi_1 = \shalf\log\lambda - \log\det R_1 - \shalf(n-1)\log\left\{\|\tilde{y}\|^2 - v_{1}^2\right\}+ \log(w/(1-w))\]
denote $\log f(\gamma^{(1)}|y)$ up to an additive constant. Under the PP stopping rule, we stop if $\pi_1 < -(n-1)\log(\|\tilde{y}\|^2)/2,$ the right side being $\log f(\phi|y)$ up to the same additive constant.

Next, we also add the second index before going into a loop. To that end, let 
\[S_1 = D^{-1/2}Z^\top (Z_{i_1} - \bar{Z}_{i_1}1_n)/D_{i_1}^{1/2},\qquad \zeta_{1} = S_1\odot S_1\]
and $\omega_{1} = \sqrt{n+\lambda - \zeta_{1}}$ where the square root is computed element-wise. Let $u_{1} = ({r} - v_1S_1)\oslash \omega_{1}.$ Then,
\[i_2 = \arg\max_{i\ne i_1}\left[ -\log\det R_1 - \log \omega_{1,i} - \shalf(n-1)\log\left\{\|\tilde{y}\|^2 - v_1^2 - u_{1,i}^2 \right\} \right].\]
Set $\gamma^{(2)} = \gamma^{(1)}\cup \{i_2\},$ $b_2 = \omega_{1,i_2},$ $v_2 = (v_1,u_{1,i_2}),$ and compute $\log f(\gamma^{(2)}|y)$ up to an additive constant as
\[\pi_2 = -\shalf 2\log\lambda - \log\det R_1 - \log b_2 - \shalf(n-1)\log\left\{\|\tilde{y}\|^2 - \|v_2\|^2\right\}+ 2\log(w/(1-w)).\]
Under the PP stopping rule, we stop if $\pi_2 < \pi_1.$

For $k\geq 3,$ until stopping we 
\begin{itemize}[leftmargin=*]
 \item[--] Compute $\alpha_{k-1} = R_{k-2}^{-\top} D_{\gamma^{(k-2)}}^{-1/2}Z_{(k-2)}^\top \left(Z_{i_{k-1}} - \bar{Z}_{i_{k-1}}1_n\right)/D_{i_{k-1}}^{1/2},$
 where \\$Z_{(k-2)}=[Z_{i_1},Z_{i_2},\cdots,Z_{i_{k-2}}].$ The order of the columns is important because the Cholesky factor is computed according the screening path.
 
 \item[--]  The Cholesky factor of $X_{\gamma^{(k-1)}}^\top X_{\gamma^{(k-1)}} + \lambda I_{k-1},$ up to an ordering of the columns, is given by
 \[R_{k-1} = \begin{pmatrix}
              R_{k-2} & \alpha_{k-1} \\ 0 & b_{k-1}
             \end{pmatrix}
\]
\item[--] Update $\log\det R_{k-1} = \log\det R_{k-2} + \log b_{k-1}.$
\item[--] Set $\eta_{k-1} = b_{k-1}^{-1}D^{-1/2}Z^\top (X_{i_{k-1}}-X_{\gamma^{(k-2)}}D_{\gamma^{(k-2)}}^{-1/2}R_{k-2}^{-1}\alpha_{k-1}).$
\item[--] Update $\zeta_{k-1} = \zeta_{k-2} + \eta_{k-1}\odot\eta_{k-1}$  and set $\omega_{k-1} = \sqrt{n + \lambda - \zeta_{k-1}}.$
\item[--] Update $u_{k-1} = (u_{k-2}\odot \omega - u_{k-2,i_{k-1}}\eta_{k-1})\oslash \omega_{k-1}.$
\item[--] Set $\gamma^{(k)} = \gamma^{(k-1)}\cup\{i_k\},$ $b_k = \omega_{k-1,i_k},$ $v_k = (v_{k-1},u_{k-1,i_k}),$ where
\[i_{k} = \arg\max_{i\notin\gamma^{(k-1)}}\left[ - \log\omega_{k-1,i} -\shalf(n-1)\log\left\{\|\tilde{y}\|^2 - \|v_{k-1}\|^2 - u_{k-1,i}\right\}\right]\]
\item[--] Compute $\log f(\gamma^{(k)}|y)$ up to an additive constant as
\[\pi_k = -\frac{k}{2}\log\lambda - \log\det R_{k-1} - \log b_k - \frac{n-1}{2}\log\left\{\|\tilde{y}\|^2 - \|v_k\|^2\right\} + k\log\frac{w}{1-w}.\]
\end{itemize}
As before, under the PP stopping criterion, we stop at the $k$th iteration and return $\gamma^{(k-1)}$ if $\pi_k < \pi_{k-1}.$

Overall, the computational complexity in the $k$th iteration is $O(k^2 + kn + np).$ Assuming the worse case scenario when the number of iterations is $O(n),$ the total computational cost is $O(n^3 + n^2p).$ If $Z$ is sparse, then this reduces to $O(n^3 + n\|Z\|_0)$ where $\|Z\|_0$ is the number of non-zero entries in $Z.$ This is same as the computational complexity of HOLP as it computes $\widehat{\beta}_{HOLP} = X^{\top}(XX^\top)^{-1}y,$ where computing $XX^\top$ incurs a cost of $O(n^2p)$ and computing $(XX^\top)^{-1}y$ incurs a cost of $O(n^3).$ Also the computational  complexity of robust rank correlation screening \citep{li:peng:zhan:zhu:2012} is $O(n^2p).$ In contrast, the computational complexities of iterated sure
independence screening \citep{fan:lv:2008} and tilting procedures \citep{cho:fryz:2012, lin:pang:2014} are much higher. Furthermore, the memory requirement of BITS, in addition to storing the original matrix $Z,$ is $O(n^2),$ mainly for storing the Cholesky factors $R_k'$s. This is same as the memory requirement of HOLP even if the matrix $X$ is not explicitly stored. The complexity of the FR method as implemented in the github repository `screening'\footnote[1]{https://github.com/wwrechard/screening} by \cite{wang:leng:2016} in the $k$th iteration is $O([k^3+k^2n]p)$ although a faster implementation of FR can be achieved by the delayed Cholesky update proposed here.

\section{Simulation studies}
\label{sec:exam}

\label{sec:posteriormass}
In this section, we study how much posterior mass is covered by BITS under 
different choices of $\lambda$ and stopping criteria. We consider our numerical study in the context of eight simulation
models described below. For these examples \ref{case:ind}--\ref{case:spurious}, the rows of $X$ are
generated from multivariate normal distributions with mean zero and different covariance matrices. In all settings except \ref{case:sparsefac}, the true model is taken to be $t=\{1,2,\ldots,9\}.$ For $i\in t$ $\beta_i$'s $\stackrel{iid}{\sim} {\cal N}(0,\sigma^2/\lambda_0)$ where $\sigma^2 = \lambda_0 = 1;$ $\beta_i = 0$ for $i \ne t.$

\begin{enumerate}[label=E.\arabic*,leftmargin=*]
 \item \label{case:ind} \textbf{Independent predictors (Ind.)} ~~Here the 
covariance matrix is $I_p.$

\item \label{case:comp}\textbf{Compound symmetry (CS)} ~~In this example, the 
covariance matrix is $\rho 11^T + (1-\rho)I_p.$. The value of $\rho$ is set to 
be equal to $0.5$.

\item \label{case:auto} \textbf{Autoregressive correlation} ~~
The auto regression, correlation structure among covariates is
appropriate when there is an ordering (say, based on time) in
covariates, and variables further apart are less correlated. We use the AR(1) structure where the $(i,j)$th entry of the covariance matrix is is $\rho^{|i-j|}.$ We set $\rho = 0.5$.

\item \label{case:fac} \textbf{Factor models (Fac.)} ~~
This example is from \cite{mein:buhl:2006} and
\cite{wang:leng:2016}. Fix $k=10.$ Let $F$ be a $p\times k$ matrix whose entries are iid $\sN(0,1).$ The covariance matrix is $FF^\top + I_p.$

\item \label{case:gr} \textbf{Group structure (Grp.)} ~~
This special correlation structure arises when variables are grouped
together in the sense that the variables from the same group are
highly correlated. This example is similar to example 4 of \cite{zou:hast:2005} where 9 true variables are assigned to 3 groups. We generate the predictors as
$X_{m} = z_1 + \zeta_{1, m}$, $X_{3+m} = z_2 + \zeta_{2, m}$,
$X_{6+m} = z_3 + \zeta_{3, m}$  where
$z_i \overset{iid}\sim \sN_n(0, I_n)$,
$\zeta_{i, m} \overset{iid}\sim \sN_n(0, 0.01I_n)$ and $z_i$'s and $\zeta_{i, m}$'s are independent for $1\leq i \leq 3$ and for $m=1,2,3$.

\item \label{case:ext} \textbf{Extreme correlation (Ext.)} ~~
We modify \pcite{wang:2009} challenging Example 4 to make it more complex.
Simulate $Z_i \stackrel{iid}{\sim} {\cal N}_{n} (0, I), i= 1, \dots, p$,
and $W_i \stackrel{iid}{\sim} {\cal N}_{n} (0, I), i= 1, \dots,9$. Set
$X_i = (Z_i + W_i)/\sqrt{2}, i= 1, \dots, 9$ and
$X_i = (Z_i + \sum_{i=1}^{9}W_i)/2$ for $i= 10, \dots, p$. The
marginal correlation between the response and any unimportant
variable is ($2.5/\sqrt{3}$ times) larger in magnitude than the same between the response and the true predictors. 

\item \label{case:sparsefac} \textbf{Sparse factor models (Sp.Fac.)} ~~
This example is a sparse version of \ref{case:fac}. Let $f_{ij}$ denotes the 
$(i,j)$th entry of $F$. Here, for each fixed $1\leq j\leq 5,$ $f_{ij} 
\sim \sN(0,1)$ if $5(j-1)+1\le i \le 5j,$ and $f_{ij} = 0$ otherwise. Also, 
$\Sigma = FF^\top + 0.01I_p$ and $\beta_j\stackrel{iid}{\sim}$ ${\cal N}(0,1)$ for $1\le j \le 25$ and 0 for $j > 25.$

\item \label{case:spurious} \textbf{Spurious correlation models (Spur.)} ~~ Here, the unimportant variables are highly correlated with the mean 
function. Specifically, $X_1,X_2,\ldots, X_9 \stackrel{iid}{\sim} {\cal 
N}_n(0,I),$ $\mu = \sum_{i=1}^9\beta_i X_i,$ and $X_i \stackrel{iid}{\sim}$ 
$\mathcal{N}_n(\mu,0.25 I)$ for $i\geq 9$.
\end{enumerate}

For each of these cases, we generate data using $(n,p) = (50,100),$ $(75, 200),$ 
$(100,2000),$ $(150,2000),$ and $(200,10000).$ We then run BITS with three different values of $\lambda:$ a large value $\lambda = p/n$ (BITS1), a moderate value $\lambda = n\log n/p$ (BITS2)  and a small value $\lambda = n/p$ (BITS3). In addition, BITS1, BITS2, BITS3 are 
run with two different stopping rules: top $n$ variables (denoted by `(n)'), 
and the PP criterion (denoted by `(PP)') using $w = |t|/p$ where $|t|$ is the  true model size. We also consider the union of the models from the three BITS  settings (UBITS) for each stopping rule.

Next, for each screened model, we use SVEN \citep{li:dutt:roy:2023} implemented in the package `bravo' \citep{bravo:2021} to find the models with significant posterior probabilities under the data generating $\lambda_0$. We also use SVEN without screening to find models with significantly high posterior probabilities. This allows us to approximate the posterior mass covered by submodels of the screened models. We compute the three metrics: true positive rate (TPR) is the percentage of true variables included in the screened model, posterior mass coverage (PMC) is an approximation to the total mass of the submodels of the screened model, and the size of the screened model. We repeat the process 100 times for each setting. We report the mean TPRs, mean PMCs, and median model sizes. For comparison,  also compute the TPRs of the three popular screening methods: HOLP \citep{wang:leng:2016}, SIS \citep{fan:lv:2008}, and forward regression \citep{wang:2009}, each with two stopping rules: models of size $n$ ($n-2$ for FR) and using EBIC. For brevity, we report here two settings: $(n,p) = (50, 100)$ and $(200,10000)$ in Tables \ref{tab.n50p100} and \ref{tab.n200p10000}; remaining cases are reported in Section \ref{sec:simufurther} of the supplementary document.

From the results, we see that the optimal value of the tuning parameter depends on the correlation structure of the covariate matrix. A large value of $\lambda$ (BITS1) works well in most of the settings, but a small value of $\lambda$ (BITS3) works better for the factor model setting, and a moderate value of $\lambda$ (BITS2) works better in the sparse factor model setting. The PP stopping criteria have substantially worse TPR than models of size $n$. However, it misses variables with small coefficients, losing only a bit of PMC while giving substantially small screened models. The union of the models from BITS always performs better than the individual BITS algorithms. Thus, UBITS should provide a good hedging against the unknown correlation among the covariates.

Very interestingly, note that the union models have smaller sizes than the sums of the sizes of the three BITS models but have substantially larger posterior mass coverage than the individual BITS. This suggests that the three BITS models overlap substantially, but each picks up some important variables not picked up by others. In order to understand this better, we focus on the compound symmetry setting with $n = 50$ and $p = 100$ where the contrast is most stark. In Table \ref{tab.effectsizevsbits}, we bin the coefficients and report how often they are picked up to be among the top $50$ variables by 
each of the three BITS settings. We see that small effect sizes are picked up more often by BITS3 with small shrinkage, whereas larger effect sizes are picked up slightly more often by BITS1, followed by BITS2 with larger shrinkage than BITS1. These findings also confirm that UBITS provides good hedging against unknown effect sizes.

\begin{table}[!htp]
\centering
\caption{Simulation results for $n = 50, p = 100.$}
{\small 
\begin{tabular}{lrrrrrrrr}
  \hline
Method & Ind. & CS & AR & Fac. & Grp. & Ext. & Sp.Fac. & Spur. \\ 
  \hline
\multicolumn{9}{l}{Mean true positive rates}\\
BITS1(n) & 84.0 & 80.8 & 87.3 & 84.2 & 92.1 & 77.6 & 83.1 & 11.6 \\ 
  BITS1(PP) & 71.9 & 64.3 & 67.3 & 58.4 & 70.2 & 74.9 & 50.4 & 7.8 \\ 
  BITS2(n) & 84.1 & 80.3 & 87.2 & 84.0 & 92.4 & 77.4 & 83.5 & 11.7 \\ 
  BITS2(PP) & 72.0 & 64.2 & 67.1 & 58.3 & 70.3 & 75.0 & 50.4 & 7.8 \\ 
  BITS3(n) & 83.9 & 80.3 & 86.6 & 87.2 & 93.7 & 77.0 & 79.8 & 9.6 \\ 
  BITS3(PP) & 72.4 & 62.9 & 64.8 & 59.7 & 52.3 & 73.9 & 49.3 & 5.0 \\ 
  UBITS(n) & 89.8 & 86.3 & 92.8 & 90.6 & 96.0 & 79.7 & 89.6 & 14.0 \\ 
  UBITS(PP) & 74.0 & 68.3 & 70.0 & 67.6 & 73.2 & 76.7 & 57.2 & 9.7 \\ 
  HOLP(n) & 79.2 & 79.6 & 80.2 & 78.6 & 83.8 & 80.9 & 71.6 & 10.3 \\ 
  HOLP(eBIC) & 34.6 & 27.1 & 34.2 & 47.3 & 31.8 & 40.0 & 15.0 & 0.3 \\ 
  SIS(n) & 77.4 & 61.9 & 80.6 & 60.6 & 90.1 & 60.0 & 71.4 & 0.0 \\ 
  SIS(eBIC) & 30.0 & 14.2 & 25.3 & 14.8 & 28.8 & 29.7 & 7.6 & 0.0 \\ 
  FR(n-2) & 82.8 & 76.9 & 79.0 & 85.9 & 58.9 & 87.2 & 65.2 & 45.9 \\ 
  FR(eBIC) & 50.2 & 36.2 & 41.4 & 47.1 & 24.6 & 47.6 & 16.9 & 0.0 \\ 
   \hline
\multicolumn{9}{l}{Mean posterior mass coverages}\\
BITS1(n) & 68.8 & 68.0 & 69.4 & 55.8 & 75.6 & 66.5 & 54.7 & 63.9 \\ 
  BITS1(PP) & 62.3 & 59.0 & 62.3 & 47.2 & 64.4 & 57.4 & 38.8 & 47.5 \\ 
  BITS2(n) & 68.7 & 67.5 & 69.1 & 56.6 & 75.3 & 65.9 & 57.0 & 63.8 \\ 
  BITS2(PP) & 62.1 & 58.7 & 62.4 & 49.3 & 64.6 & 56.8 & 39.8 & 47.4 \\ 
  BITS3(n) & 68.2 & 66.6 & 67.7 & 59.9 & 73.5 & 64.1 & 54.6 & 60.0 \\ 
  BITS3(PP) & 61.6 & 57.9 & 61.1 & 51.9 & 63.1 & 55.4 & 39.0 & 45.1 \\ 
  UBITS(n) & 91.1 & 94.4 & 91.2 & 88.1 & 97.8 & 90.3 & 97.6 & 99.0 \\ 
  UBITS(PP) & 85.3 & 86.1 & 85.4 & 80.2 & 90.0 & 81.4 & 84.1 & 86.1 \\ 
   \hline
\multicolumn{9}{l}{Median model sizes}\\
BITS1(PP) & 14.0 & 15.0 & 12.0 & 12.0 & 15.0 & 10.0 & 17.5 & 22.0 \\ 
  BITS2(PP) & 14.0 & 15.0 & 12.0 & 12.0 & 15.0 & 10.5 & 17.0 & 21.5 \\ 
  BITS3(PP) & 14.5 & 15.5 & 11.5 & 11.0 & 10.0 & 9.5 & 19.0 & 19.0 \\ 
  UBITS(n) & 67.0 & 65.0 & 66.0 & 62.0 & 66.0 & 60.0 & 65.0 & 58.0 \\ 
  UBITS(PP) & 17.0 & 21.0 & 15.0 & 16.0 & 17.0 & 15.0 & 22.5 & 27.0 \\ 
  HOLP(eBIC) & 3.0 & 2.0 & 3.0 & 5.0 & 3.0 & 3.5 & 2.5 & 1.0 \\ 
  SIS(eBIC) & 2.0 & 1.0 & 2.0 & 2.0 & 1.0 & 2.0 & 1.0 & 1.0 \\ 
  FR(eBIC) & 5.0 & 3.0 & 4.0 & 8.0 & 2.0 & 5.0 & 4.0 & 1.0 \\ 
   \hline
\end{tabular}
}
\label{tab.n50p100}
\end{table}

\begin{table}[!htp]
\centering
\caption{Simulation results for $n = 200, p = 10,000.$}
{\small 
\begin{tabular}{lrrrrrrrr}
  \hline
Method & Ind. & CS & AR & Fac. & Grp. & Ext. & Sp.Fac. & Spur. \\ 
  \hline
\multicolumn{9}{l}{Mean true positive rates}\\
BITS1(n) & 73.0 & 69.6 & 83.3 & 61.7 & 89.7 & 84.0 & 68.9 & 9.0 \\ 
  BITS1(PP) & 73.0 & 69.6 & 81.8 & 61.7 & 88.7 & 84.0 & 65.5 & 9.0 \\ 
  BITS2(n) & 76.4 & 71.6 & 84.9 & 80.1 & 91.0 & 83.7 & 85.4 & 0.1 \\ 
  BITS2(PP) & 76.2 & 70.9 & 74.0 & 77.7 & 49.0 & 83.3 & 75.5 & 0.1 \\ 
  BITS3(n) & 76.6 & 71.3 & 82.4 & 79.3 & 91.3 & 83.7 & 86.2 & 0.2 \\ 
  BITS3(PP) & 76.2 & 70.8 & 74.0 & 75.2 & 38.6 & 83.3 & 75.6 & 0.2 \\ 
  UBITS(n) & 78.2 & 74.3 & 90.6 & 80.4 & 92.0 & 87.1 & 89.3 & 9.2 \\ 
  UBITS(PP) & 78.1 & 74.0 & 86.4 & 80.6 & 89.7 & 86.9 & 83.1 & 9.3 \\ 
  HOLP(n) & 61.6 & 64.1 & 72.6 & 56.8 & 83.0 & 87.3 & 52.8 & 48.2 \\ 
  HOLP(eBIC) & 45.9 & 42.6 & 41.4 & 37.1 & 43.7 & 69.4 & 21.0 & 23.2 \\ 
  SIS(n) & 61.4 & 45.1 & 72.8 & 14.4 & 82.8 & 54.0 & 53.0 & 0.0 \\ 
  SIS(eBIC) & 45.8 & 22.9 & 41.6 & 2.4 & 43.9 & 46.9 & 20.7 & 0.0 \\ 
  FR(n-2) & 76.7 & 71.0 & 74.0 & 79.9 & 33.9 & 84.2 & 75.6 & 1.8 \\ 
  FR(eBIC) & 67.9 & 63.4 & 65.4 & 68.4 & 27.8 & 70.8 & 67.2 & 0.0 \\ 
   \hline
\multicolumn{9}{l}{Mean posterior mass coverages}\\
BITS1(n) & 55.6 & 44.6 & 53.2 & 2.7 & 81.3 & 78.9 & 2.8 & 82.9 \\ 
  BITS1(PP) & 55.2 & 44.0 & 52.7 & 2.7 & 79.2 & 77.0 & 2.8 & 72.0 \\ 
  BITS2(n) & 75.9 & 66.6 & 75.2 & 43.0 & 82.4 & 80.5 & 70.3 & 38.7 \\ 
  BITS2(PP) & 75.1 & 65.8 & 74.3 & 41.7 & 80.8 & 77.7 & 68.3 & 36.4 \\ 
  BITS3(n) & 76.0 & 66.3 & 74.5 & 40.8 & 82.4 & 78.7 & 69.9 & 29.1 \\ 
  BITS3(PP) & 75.2 & 65.4 & 73.8 & 40.6 & 80.9 & 77.1 & 68.0 & 27.9 \\ 
  UBITS(n) & 83.4 & 78.0 & 82.0 & 52.3 & 91.6 & 92.3 & 74.7 & 95.7 \\ 
  UBITS(PP) & 82.6 & 77.0 & 81.2 & 51.8 & 90.1 & 89.2 & 73.6 & 89.1 \\ 
   \hline
\multicolumn{9}{l}{Median model sizes}\\
BITS1(PP) & 136.0 & 200.0 & 125.5 & 200.0 & 79.5 & 200.0 & 116.5 & 200.0 \\ 
  BITS2(PP) & 75.0 & 76.0 & 76.0 & 64.5 & 76.0 & 9.0 & 66.0 & 70.0 \\ 
  BITS3(PP) & 88.0 & 88.0 & 88.5 & 77.0 & 86.0 & 9.0 & 79.0 & 77.0 \\ 
  UBITS(n) & 387.0 & 386.0 & 386.5 & 387.0 & 384.0 & 383.0 & 379.0 & 373.0 \\ 
  UBITS(PP) & 232.0 & 298.0 & 227.0 & 299.5 & 191.0 & 201.0 & 195.5 & 307.5 \\ 
  HOLP(eBIC) & 4.0 & 4.0 & 4.0 & 7.0 & 4.0 & 6.0 & 5.0 & 4.0 \\ 
  SIS(eBIC) & 4.0 & 2.0 & 4.0 & 3.0 & 4.0 & 5.0 & 5.0 & 1.0 \\ 
  FR(eBIC) & 6.0 & 6.0 & 6.0 & 14.0 & 3.0 & 7.0 & 18.0 & 1.0 \\ 
   \hline
\end{tabular}
}
\label{tab.n200p10000}
\end{table}

\begin{table}[htp]

\caption{Proportion of the times an important variable with coefficient $\beta$ 
is picked up by a BITS algorithm in the compound symmetry setting with $n = 
50, p = 100.$}
\label{tab.effectsizevsbits}
\centering
{\small%
\begin{tabular}{l|lllllllll}
  \hline
  $|\beta|$& (0, .05] & (.05, .1] & (.1, .15] & (.15, .2] & (.2, .25] 
& (.25, .3] & (.3, .5] & (.5, 1] & $> 1$ \\ 
  \hline
  BITS1 & 36.36 & 45.95 & 43.59 & 36.84 & 46.43 & 59.46 & 69.17 & 94.40 & 100 
\\ 
  BITS2 & 39.39 & 51.35 & 46.15 & 31.58 & 46.43 & 59.46 & 66.17 & 93.66 & 100 
\\ 
  BITS3 & 54.55 & 51.35 & 41.03 & 44.74 & 39.29 & 59.46 & 64.66 & 93.28 & 98.95 
\\ 
  \hline
\end{tabular}
}
\end{table}

BITS also appears to perform as well as or better than the frequentist screening methods. Generally, one of the three BITS settings closely matches or beats the best frequentist method in all but the spurious correlation setting. HOLP has better asymptotic TPR than the other methods under the spurious correlation structure. However, UBITS still captures substantial posterior mass because there are different models with high posterior probabilities.


\section{Real data example}
\label{sec:data}


We compare the screening methods using a real data set from \citep{cook:mcmu:2012} on a genomewide association study for maize starch, protein and oil contents. 
The original field trial at Clayton, NC in 2006 consisted of more than 5,000 inbred lines and check varieties primarily coming from a diverse panel consisting of 282 founding lines. The response from the field trials are typically spatially correlated, 
thus we use a random
row-column adjustment to obtain the adjusted phenotypes of the
varieties. However, marker information of only $n=3,951$ of these
varieties are available from the panzea project
(https://www.panzea.org/) which provides information on 546,034 single
nucleotide polymorphisms (SNP) markers after removing duplicates and
SNPs with minor allele frequency less than 5\%. We use the
starch content as our phenotype for conducting the association
study. Because the inbred varieties are bi-allelic, we store the
marker information in a sparse format by coding the minor alleles by
one and major alleles by zero.

\begin{figure}[htbp]
\begin{center}
\includegraphics[width = \textwidth]{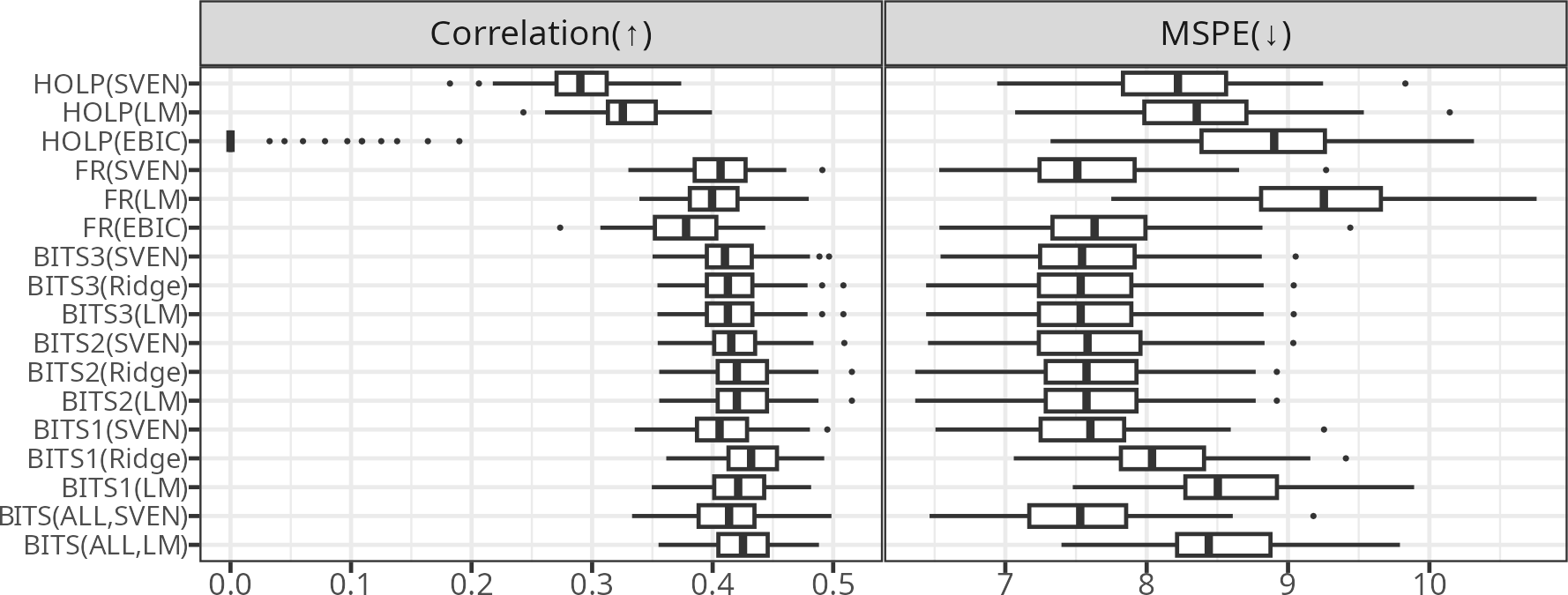}
~ 

~
\begin{tabular}{ccccccc}
Method    & BITS1       & BITS2       & BITS3       & BITS(ALL)        & HOLP(EBIC)  & FR(EBIC)     \\
\hline
Mean (SE) & 182.9 (16.3) & 46.5 (6.4) & 32.1 (5.6) & 213.2 (18.0) & 1.18 (0.47) & 8.96 (1.09)
\end{tabular}
\end{center}
    \caption{Prediction accuracy on the test sets: correlation (top left), MSPE (top right) using least squares (LM), ridge estimates (Ridge), and Bayesian model averaging (SVEN). The means and standard deviations of model sizes are shown in the table.}
     \label{fig:realdata}
\end{figure}

In this study, we do a cross-validation by randomly splitting the whole data set with training set of size $n= 3,200$ and testing set of size 751. We run BITS with PP stopping rule with $w$ fixed at  $0.1$ but with three distinct values of $\lambda$: $p/n$ (BITS1), $n\log(n)/p$ (BITS2) and $n/p$ (BITS3). We also consider the union of these models. We compare these methods with HOLP and FR. The EBIC stopping criterion is also applied to all these methods for comparison. We repeat the process 100 times. In order to be able to use least squares estimates of the regression coefficients, we do not use screened model size same as $n.$ However, HOLP and FR result in very small screened models (Figure \ref{fig:realdata}) under EBIC based stopping rule. Indeed, out of 100 repetitions, 44 times HOLP with EBIC results in a null model. 

In order to keep the comparisons fair, we also use HOLP and FR with same model size as the union of models from the BITS1, BITS2 and BITS3. For each of the three BITS methods, we use both the ordinary least squares estimates of the regression coefficients (denoted by `LM') and the ridge estimates(denoted by `Ridge') with the associated $\lambda$ because they are the posterior mode of the regression coefficient given the screened model (Figure~\ref{fig:realdata}). In addition, we perform a refined variable selection using SVEN \citep{li:dutt:roy:2023} for each of these screened models (denoted by `SVEN'). SVEN allows prediction through Bayesian model averaging. 

From the boxplots in Figures~\ref{fig:realdata}, it can be seen that models selected by BITS and FR with SVEN stopping rule yield very good prediction accuracies. In particular, union of BITS variables together with Bayesian model averaging yield more accurate MSPE than other BITS methods on an average. HOLP with EBIC stopping rule appears to be overtly conservative. Overall, HOLP with either EBIC or SVEN results in lowest prediction accuracies.



\section{Discussion}\label{sec:disc}


In this paper, we propose a Bayesian iterative screening (BITS) method for screening variables in an ultra-high dimensional regression model that can accommodate prior information on the effect size and the model size. Despite being built on a Gaussian model assumption, BITS has been shown to be screening consistent even when the family is misspecified. In contrast to SIS, BITS does not require strong assumptions on marginal correlations. Compared to the frequentist iterative screening method FR, BITS naturally accommodates penalization on the effect size enhancing screening accuracy, particularly when important predictors are correlated among themselves. The proposed PP stopping rule, which is shown to be screening consistent, can provide informative screening by incorporating prior knowledge on the hyperparameters. BITS is implemented by a sophisticated algorithm that attains the same computational complexity as HOLP and allows fast statistical computations in ultra-high dimensional problems. 
BITS has been shown to have better performance than HOLP and FR in most simulation settings, especially when the BITS models from different shrinkages are united. Thus, in practice, it could be useful to take union of models from BITS using various degrees of shrinkage. Then Bayesian or other variable selection algorithms may find it easier to discover useful models. Finally, models from BITS are shown to have substantial posterior mass coverage in all simulation settings. 

As a future project, we plan to establish the posterior screening
consistency of BITS without the assumption of strong selection
consistency and calibrate the stopping rule to return a set of variables that constitutes a subspace of models containing at least a prespecified posterior probability. BITS can be extended to accommodate larger class of
models, for example generalized linear regression models. In theory,
the iterative algorithm \eqref{eqn:bits-general} is quite general and
does not require Gaussianity assumption. We may use Laplace
approximations to achieve fast statistical computations in generalized
linear
models. 
Also, our future projects include developing Bayesian screening
methods for non-linear, partial linear and functional linear models.

\begin{supplement}
  \stitle{Some useful notations and results}
  \sdescription{This section contains notations used in the proofs and some results.}
\end{supplement}
\begin{supplement}
\stitle{Proofs of theorems}
\sdescription{Proofs of the theorems stated in the paper appear here.}
\end{supplement}
\begin{supplement}
\stitle{Proofs of lemmas and corollaries}
\sdescription{This section contains proofs of several lemmas and some other results.}
\end{supplement}
\begin{supplement}
  \stitle{Further simulation results} \sdescription{Simulation results
    for further values of $n$ and $p$.}
\end{supplement}

\section*{Acknowledgement}
We thank the Editor, an anonymous Associate Editor, and two anonymous reviewers for their helpful comments and suggestions which led to an improved version of the manuscript. The work was partially supported by USDA NIFA 2023-70412-41087 (S.D., V.R.), USDA Hatch Project IOW03717 (S.D.), and the Plant Science Institute (S.D.).

\bibliographystyle{ba}
\bibliography{ref}

\newpage
\pagebreak

\markright{ \hbox{\footnotesize\rm Statistica Sinica: Supplement
}\hfill\\[-13pt]
\hbox{\footnotesize\rm
}\hfill }

\markboth{\hfill{\footnotesize\rm Run Wang, Somak Dutta, and Vivekananda Roy} \hfill}
{\hfill {\footnotesize\rm Bayesian iterative screening} \hfill}

\renewcommand{\thefootnote}{}
$\ $\par \fontsize{12}{14pt plus.8pt minus .6pt}\selectfont


 \centerline{\large\bf Bayesian iterative screening in ultra-high dimensional}
\vspace{2pt}
 \centerline{\large\bf linear regressions}
 \vspace{.25cm}
 \centerline{Run Wang, Somak Dutta, and Vivekananda Roy} 
\vspace{.4cm} 
\centerline{\it Department of Statistics, Iowa
   State University, USA}
\vspace{.55cm}
 \centerline{\bf Supplementary Material}
\vspace{.55cm}
\fontsize{9}{11.5pt plus.8pt minus .6pt}\selectfont
\par

\def\theequation{S\arabic{section}.\arabic{equation}}
\def\thesection{S\arabic{section}}

\fontsize{12}{14pt plus.8pt minus .6pt}\selectfont

\setcounter{equation}{0}
\setcounter{figure}{0}
\setcounter{table}{0}
\setcounter{page}{1}
\setcounter{section}{0}
\makeatletter
\renewcommand{\thesection}{S\arabic{section}}
\renewcommand{\thesubsection}{\thesection.\arabic{subsection}}
\renewcommand{\theequation}{S\arabic{equation}}
\renewcommand{\thefigure}{S\arabic{figure}}
\renewcommand{\bibnumfmt}[1]{[S#1]}
\renewcommand{\citenumfont}[1]{S#1}
\renewcommand{\thetable}{S\arabic{table}}

\section{Some useful notations and results}\label{sec:notationsandresults}
We use the following notations:
\begin{itemize}
    \item $\tau^+$ denotes the largest eigenvalue of $X_\gamma^\top X_\gamma/n$ for all $|\gamma|\leq K_n|t|.$
    \item $\Phi$ and $\phi$ denote the standard normal distribution and density function, respectively.
\end{itemize}
\noindent For any model $\gamma$ with $|\gamma| < n,$
\begin{itemize}
    \item $\tilde{\beta}_\gamma = (X_\gamma^\top X_\gamma+\lambda I)^{-1}X_\gamma^\top \tilde{y}$ and $H_\gamma = X_\gamma (X_\gamma^\top X_\gamma)^{-1}X_\gamma^\top.$
    \item $\rssl(\gamma) = \tilde{{y}}^\top\tilde{{y}} - \tilde{{y}}^\top{X}_{\gamma}({X}_{\gamma}^\top{X}_{\gamma }+ \lambda{I})^{-1}{X}_{{\gamma}}^\top\tilde{{y}}.$
    \item $\rss(\gamma) = \tilde{{y}}^\top\tilde{{y}} - \tilde{{y}}^\top{X}_{\gamma}({X}_{\gamma}^\top{X}_{\gamma } )^{-1}{X}_{{\gamma}}^\top\tilde{{y}}$ = $\tilde{{y}}^\top(I-H_\gamma)\tilde{{y}}.$
    \item 
    $\Omega(\gamma) =  -\dfrac{|\gamma|}{2(n-1)}\log\lambda + \dfrac{1}{2(n-1)}\log|X_\gamma^\top X_\gamma + \lambda I| + \frac12\log RSS_\lambda(\gamma) -$ \\  
    $\dfrac{|\gamma|}{n-1}\log w - \dfrac{p-|\gamma|}{n-1}\log(1-w)$
which is equal to $-\log f(\gamma|y)/(n-1)$ up to an additive constant that does not depend on $\gamma.$
\end{itemize}

First we state some useful results whose proofs are given in Section \ref{sec:supp_proofs} of the supplement. These results are used in the proofs of Theorems \ref{thm:consistency} and \ref{thm.gen_stop}.

\begin{lemma}\label{lem.rssdiff}
 For any model $\gamma$  with $|\gamma| < n,$ and any $i\notin \gamma,$
 \[\rss(\gamma) - \rss(\gamma+e_i) = \dfrac{\left\{X_i^\top\tilde{y} - X_i^\top H_\gamma\tilde{y}\right\}^2}{n - X_i^\top H_\gamma X_i}.\]
\end{lemma}

\begin{lemma}\label{lem.logdetdiffbound}
 For any model $\gamma$ and $j\notin \gamma,$
 $\log|X_{\gamma'}^\top X_{\gamma'} + \lambda I| - \log|X_{\gamma}^\top X_{\gamma}+\lambda I|$ lies between $\log\lambda$ and $\log(n+\lambda)$ where $\gamma' = \gamma + e_j.$
\end{lemma}

\begin{lemma}\label{lem.constants}
 Under conditions C2 and C3, there exists $c' > 0$ and $c'' > 0$ such that for all sufficiently large $n,$ 
\begin{equation}\label{eq.cprime}
 \log(1 + nK_n|t|/\lambda) \le 2c'\log n, \textrm{ and}
\end{equation}
\begin{equation}\label{eq.cdoubleprime}
\dfrac{K_n\lambda}{2n\tau_+} + \dfrac{K_n\log(n+\lambda)}{2(n-1)} - \dfrac{K_n\log\lambda}{2(n-1)} \le c''\log n.
\end{equation}
\end{lemma}

\begin{lemma}\label{lem.omegalowerbound}
Suppose $|\gamma| <K_n|t|$ and $\gamma\nsupseteq t$ then $\max_{k\in t\backslash \gamma} \{ \Omega(\gamma) - \Omega(\gamma+e_k)\} \ge \max_{k\in t\backslash \gamma}\{\rss(\gamma) - \rss(\gamma+e_k)\}/(2\|\tilde{y}\|^2) - \lambda/(2n\tau_+) - \log(n/\lambda+1)/(2(n-1)) + \log(w/(1-w))/(n-1).$
\end{lemma}

\begin{lemma}\label{lem.regssbound}
Suppose $|\gamma| < K_n|t|$ and $\gamma \nsupseteq t,$ then with $s_n$ given in C3,
\[\max_{k\in t\backslash \gamma} (X_k^\top (I-H_\gamma)X_t\beta_t)^2 \geq ns_n^2.\]
\end{lemma}

\begin{corollary}\label{cor.maxmaxbound}
 Suppose $|\gamma| < K_n|t|,$ $\gamma \nsupseteq t,$ and $\sigma\Upsilon_n < s_n$. Let 
 \[a_{\gamma',l} = (I - H_{\gamma'})X_l/\|(I-H_{\gamma'})X_l\| \textrm{ and } \Upsilon_n = \max_{l\in t}\max_{\gamma' : |\gamma'| < K_n|t|} |a_{\gamma',l}^\top \epsilon|.
\]
Then $\max_{k\in t\backslash\gamma} \{\rss(\gamma) - \rss(\gamma + e_k)\} \geq (s_n - \sigma\Upsilon_n)^2.$
\end{corollary}

\section{Proofs of theorems}

\begin{proof}[Proof of Theorem \ref{thm:orthogonal}]
Under the orthogonal design,
\[f(\gamma + e_i|y) > f(\gamma + e_j|y) \iff (X_i^\top \tilde{y})^2 > (X_j^\top \tilde{y})^2,\]
which does not depend on $\gamma.$ Now, for each $i \in t,$ suppose $A_i$ denotes the event
$\{(X_i^\top \tilde{y})^2 > \max_{j\notin t}(X_j^\top \tilde{y})^2\}.$ Then the theorem will be proved if $P\left(\cap_{i\in t}A_i\right) \to 1$ as $n\to\infty,$ for which it is sufficient to show that $\sum_{i\in t}P(A_i^c) \to 0,$ as $n\to\infty.$ To that end, assume without loss that true $\sigma^2 = 1 $ and notice that for $1\le l \le p,$ $U_l := n^{-1/2}X_l^\top \epsilon$ are iid standard normal because $X^\top X = nI_p.$ Thus for all $i \in t$ and $j \notin t,$ $X^\top 1 = 0$ implies that
$X_i^\top\tilde{y} = n\beta_i + \sqrt{n}U_i$ and $X_j^\top\tilde{y} = \sqrt{n}U_j.$
Thus, $A_i = \{ (\sqrt{n}\beta_i + U_i)^2 > \max_{j\notin t} U_j^2\}.$ Consequently, since $U_l$'s are iid standard normal, we have,
\begin{eqnarray}\label{eqn:PAC_bound1}
P(A_i^c) & = & 1 - P(A_i) = 1 - E\prod_{j\notin t}P(|U_j| \le |\sqrt{n}\beta_i + U_i|~\big|~U_i)\nonumber\\
& = & 1 - E(2\Phi(|\sqrt{n}\beta_i + U_i|) - 1)^{(p-|t|)} ~\le~ 1 - E(2\Phi(|\sqrt{n}\beta_i + U_i|) - 1)^{n} \nonumber \\
& = & E\left[ \left\{ 1 - (2\Phi(|\sqrt{n}\beta_i + U_i|) - 1)^{n}\right\} I(|U_i| > \sqrt{2\log n}) \right] + \nonumber \\
&  & E\left[ \left\{ 1 - (2\Phi(|\sqrt{n}\beta_i + U_i|) - 1)^{n}\right\} I(|U_i| \le \sqrt{2\log n}) \right].
\end{eqnarray}
Because $2\Phi(|\sqrt{n}\beta_i + U_i|) - 1 < 1$ and $1-\Phi(x) \le \phi(x)/x$ for all $x > 0,$  the first term on the right side of \eqref{eqn:PAC_bound1} is at most
\begin{equation}\label{eqn:PAC_bound2}
P(|U_i| > \sqrt{2\log n}) \le {2}/({n\sqrt{4\pi\log n}}).
\end{equation}

Next, since $\sqrt{n}\beta_+ \succ \sqrt{\log n},$ for all sufficiently large $n$, $\sqrt{n}\beta_+ - \sqrt{2\log n} > c_n := 2\sqrt{\log n}.$ Thus, the second term on the right side of \eqref{eqn:PAC_bound1} is at most
\begin{eqnarray}\label{eqn:PAC_bound3}
 & & E\left[ \left\{ 1 - (2\Phi(\sqrt{n}|\beta_i| -  |U_i|) - 1)^{n}\right\} I(|U_i| \le \sqrt{2\log n}) \right] \nonumber \\
 & & \le 1 - (2\Phi(\sqrt{n}\beta_+ -  \sqrt{2\log n}) - 1)^{n} 
 \le 1 - (2\Phi(c_n)-1)^n  \nonumber \\
 & & \le 2n[1-\Phi(c_n)]/c_n \le \dfrac{1}{n\log n\sqrt{8\pi}} \le \dfrac{2}{n\sqrt{8\pi\log n}},
\end{eqnarray}
where the third last inequality holds because $(1-x)^n \geq 1 - nx$, for $x \leq 1.$

Thus, from \eqref{eqn:PAC_bound1}, \eqref{eqn:PAC_bound2} and \eqref{eqn:PAC_bound3} we have, for all sufficiently large $n$, that
\[
\sum_{i \in t}P(A_i^c) \le \sum_{i\in t} \left\{\dfrac{2}{n\sqrt{4\pi\log n}} +  \dfrac{2}{n\sqrt{8\pi\log n}}\right\} = \dfrac{2(\sqrt{2}+1)|t|}{n\sqrt{8\pi\log n}}. 
\]
The right hand side $\to 0$ as $n \rightarrow \infty$ because $|t| = O(n^a)$ for some $a < 1.$
\end{proof}

\begin{proof}[Proof of Theorem \ref{thm:consistency}]
Let $\theta_0 = 0$ and $\theta_1,\theta_2,\ldots,\theta_{|t|}$ denote the random inclusion times of the variables in $t$ in the models $\gamma^{(1)},\gamma^{(2)},\ldots.$ That is, $\theta_1 = \arg\min_{i\geq 1}\{\gamma^{(i)} \cap t \ne \emptyset\}$ and for $j\geq 2,$ 
\[\theta_j = \arg\min_{i> \theta_{j-1}}\{ (\gamma^{(i)}\backslash \gamma^{(\theta_{j-1})})\cap t \ne \emptyset \}.\]
We note that,
\begin{eqnarray}\label{eq.diffbound}
\{\gamma^{(K_n|t|)} \nsupseteq t\} & = & \{\theta_{|t|} > K_n|t| \} \subseteq \bigcup_{j=0}^{|t|-1} \{\theta_{j+1} - \theta_{j} > K_n\} \nonumber\\
 & = & \biguplus_{j=0}^{|t|-1} \left\{\theta_{j+1} - \theta_{j} > K_n, \textrm{ and } \theta_{i+1} - \theta_i \le K_n~\forall 0 \le i< j\right\} \nonumber\\ 
& \subseteq & \bigcup_{j=0}^{|t|-1} \{\theta_{j+1} - \theta_j > K_n, \theta_j \le jK_n\},
\end{eqnarray}
where $\biguplus$ denotes union of disjoint sets. We now analyze the $j$th event $\{\theta_{j+1} - \theta_j > K_n, \theta_j \le jK_n\}.$ Note that, $\theta_{j+1} - \theta_j > K_n$ implies that for each $i=\theta_{j},\theta_{j}+1,\ldots,\theta_{j}+K_n - 1,$ it must be that a variable outside of $t$ was selected in the $(i+1)$st iteration, that is, for each such $i$, $\Omega(\gamma^{(i)}) - \Omega(\gamma^{(i+1)}) > \max_{k\in t\backslash\gamma^{(i)}}\left\{\Omega(\gamma^{(i)}) - \Omega(\gamma^{(i)}+e_k)\right\};$ and also that $i<K_n|t|$ and $\gamma^{(i)}\nsupseteq t.$

Thus using Lemma \ref{lem.omegalowerbound}, Corollary \ref{cor.maxmaxbound}, and \eqref{eq.cdoubleprime}, we get
\begin{eqnarray}\label{eq.lowerbound}
  & & \qquad \{\theta_{j+1} - \theta_{j} > K_n, \theta_j \le jK_n \}\cap \{\sigma\Upsilon_n < s_n\} \\
 &  & \qquad \subseteq  \bigg\{\sum_{i=\theta_j}^{\theta_j+K_n-1}\left(\Omega(\gamma^{(i)}) - \Omega(\gamma^{(i+1)})\right) \ge 
 \sum_{i=\theta_j}^{\theta_j+K_n-1}\max_{k\in t\backslash\gamma^{(i)}}\left(\Omega(\gamma^{(i)}) - \Omega(\gamma^{(i)}+e_k)\right) 
\bigg\}\nonumber\\
     & & \subseteq \bigg\{\sum_{i=\theta_j}^{\theta_j+K_n-1}\left(\Omega(\gamma^{(i)}) - \Omega(\gamma^{(i+1)})\right) \ge 
 \frac{K_n}{2\|\tilde{y}\|^2}\left|s_n- \sigma  \Upsilon_n \right|^2 - c''\log n + \frac{K_n}{n-1}\log\dfrac{w}{1-w}
\bigg\} \nonumber.
\end{eqnarray}

However, on the other hand, on $\{\theta_{j+1} - \theta_j > K_n, \theta_j \le jK_n\}$ we have using Lemma \ref{lem.logdetdiffbound}
\begin{eqnarray}
\label{eq:telesum}
& & \sum_{i=\theta_j}^{\theta_j+K_n-1}(\Omega(\gamma^{(i)}) - \Omega(\gamma^{(i+1)})) \le \dfrac{1}{2}\sum_{i=\theta_j}^{\theta_j+K_n-1}
    (\log(\rssl(\gamma^{(i)}))-\log(\rssl(\gamma^{(i+1)}))) + \nonumber\\
& & \qquad \qquad \qquad \qquad \qquad\qquad   \sum_{i=\theta_j}^{\theta_j+K_n-1}\left\{\dfrac{\log\lambda}{2(n-1)} - \dfrac{\log\lambda}{2(n-1)} + \dfrac{1}{n-1}\log\dfrac{w}{1-w}\right\}\nonumber \\
&& \qquad\qquad \qquad = \dfrac{1}{2} \left\{\log(\rssl(\gamma^{(\theta_j)}))-\log(\rssl(\gamma^{(\theta_j+K_n)})) \right\}+ \dfrac{K_n}{n-1}\log\dfrac{w}{1-w}\nonumber\\
&& \qquad\qquad \qquad<\dfrac{1}{2} \left\{ \log\|\tilde{y}\|^2-\log(\rssl(\gamma^{(\theta_j+K_n)}))\right\} + \dfrac{K_n}{n-1}\log\dfrac{w}{1-w} \nonumber\\
& & \qquad \qquad \qquad\le \dfrac{1}{2} \log(1+nK_n|t|/\lambda) + \dfrac{K_n}{n-1}\log\dfrac{w}{1-w}
\end{eqnarray}
because $\tau^+ \le K_n|t|$ and 
\begin{eqnarray*}
\rssl(\gamma^{(\theta_j+K_n)})  & = & \tilde{y}^\top \left(I+\frac{1}{\lambda}X_{\gamma^({\theta_j+K_n})} X_{\gamma^({\theta_j+K_n})}^\top\right)^{-1}\tilde{y} \geq \|\tilde{y}\|^2(1+n\tau^+/\lambda)^{-1} \\
& \geq & \|\tilde{y}\|^2(1+nK_n|t|/\lambda)^{-1}. 
\end{eqnarray*}
Thus applying \eqref{eq.cprime} from \eqref{eq:telesum} we get
\begin{equation}\label{eq.upperbound}
\sum_{i=\theta_j}^{\theta_j+K_n-1}(\Omega(\gamma^{(i)}) - \Omega(\gamma^{(i+1)})) \le c'\log n + \dfrac{K_n}{n-1}\log\dfrac{w}{1-w}.
\end{equation}

Hence, with $c = 2(c' +c''),$ using \eqref{eq.lowerbound}, \eqref{eq.upperbound} and condition C3, for sufficiently large $n,$ for all $j\le|t|,$
\begin{eqnarray*}
& & \{\theta_{j+1} - \theta_j > K_n, \theta_j \le jK_n\} \cap \{\sigma \Upsilon_n < s_n\}\\
& \subseteq & \left\{
c'\log n \geq 
\frac{K_n}{2\|\tilde{y}\|^2}(s_n -  \sigma \Upsilon_n)^2 - c''\log n\right\} \cap \{\sigma \Upsilon_n < s_n\}\\
& \subseteq & \left\{
\frac{K_n}{\|\tilde{y}\|^2}(s_n -  \sigma \Upsilon_n )^2 \le c\log n \right\} \cap \{\sigma \Upsilon_n < s_n\} \subseteq  \left\{
 s_n\left(1 - \sqrt{{\|\tilde{y}\|^2}/({nu_n})}\right)
\le \sigma \Upsilon_n < s_n \right\}
\end{eqnarray*}
where $u_n = \tau_+^2\beta_+^4K_n/(c|t|\|\beta_t\|^2\log n).$ Thus,
\begin{eqnarray*}
\{\theta_{j+1} - \theta_j > K_n, \theta_j \le jK_n\} & \subseteq& \{\{\theta_{j+1} - \theta_j > K_n, \theta_j \le jK_n\}  \cap \{\sigma \Upsilon_n < s_n\}\} \cup \{\sigma \Upsilon_n \ge s_n\} \\
& \subseteq & \left\{
 s_n\left(1 - \sqrt{{\|\tilde{y}\|^2}/({nu_n})}\right)
\le \sigma \Upsilon_n < s_n \right\} \cup \{\sigma \Upsilon_n \ge s_n\} \\
 & = & \left\{
 s_n\left(1 - \sqrt{{\|\tilde{y}\|^2}/({nu_n})}\right)
\le \sigma \Upsilon_n \right\}. 
\end{eqnarray*} 

Since for all sufficiently large $n,$ the above is true for all $j,$ we have from \eqref{eq.diffbound} that 
\begin{equation}\label{eq.finalbound}
\{\theta_{|t|} > K_n|t|\} \subseteq \left\{
\sigma \Upsilon_n \geq s_n\left(1 - \sqrt{{\|\tilde{y}\|^2}/({nu_n})}\right)
 \right\}.
\end{equation}
Thus, for all sufficiently large $n,$ using the union bound,
\begin{eqnarray*}
& & P(\theta_{|t|} > K_n|t|) \le  P\left(
\sigma \Upsilon_n \geq s_n\left(1 - \sqrt{{\|\tilde{y}\|^2}/({nu_n})}\right)
 \right) \\
&  & \quad \leq P\left(\sigma \Upsilon_n \geq s_n\left(1 - \sqrt{{\|\tilde{y}\|^2}/({nu_n})}\right), \|\tilde{y}\|^2 \le nu_n (1-\sigma \delta)^2\right) +  P(\|\tilde{y}\|^2 > nu_n (1-\sigma \delta)^2)\\
&  & \quad \leq  P(\Upsilon_n \geq s_n\delta) + P(\|\tilde{y}\|^2 > nu_n (1-\sigma \delta)^2) \\
&  & \quad \leq \sum_{l\in t}\sum_{\gamma:|\gamma|<K_n|t|}P\left(|a_{\gamma,l}^\top \epsilon| > s_n\delta\right)  + P(\|\tilde{y}\|^2 > nu_n (1-\sigma \delta)^2) \\
&  & \quad \leq \sum_{l\in t}\sum_{\gamma:|\gamma|<K_n|t|}e^{1-q(s_n\delta)} + P(\|\tilde{y}\|^2 > nu_n (1-\sigma \delta)^2) \\
&  & \quad \leq \exp(1 - q(s_n\delta) + K_n|t|\log p + \log|t|) +  P(\|\tilde{y}\|^2 > nu_n (1-\sigma \delta)^2).
\end{eqnarray*}
This proves the theorem.
\end{proof}

\begin{proof}[Proof of Theorem \ref{thm.ortho_stop}]
Assume without loss that the true $\sigma^2 = 1.$ First we shall show that $P(\mathcal{T} \geq |t|) \to 1.$ 
As in the proof of Theorem \ref{thm:orthogonal} note that for any $i\in t, (X_i^\top \tilde{y})^2 = (n\beta_i + \sqrt{n}U_i)^2$ where $U_l$'s are i.i.d $N(0,1)$ variables. Thus for any $\gamma \subset t,$ and $i\in t \cap \gamma^c,$ $f(\gamma + e_i |y) < f(\gamma|y)$ is equivalent to,
\begin{eqnarray}
\label{proof:maxz1}
& & \log a_n + (n-1)\log(RSS_\lambda(\gamma)) < (n-1)\left\{\log\left(RSS_\lambda(\gamma) - {(X_i^\top \tilde{y})^2}/{(n+\lambda)}\right)\right\}\nonumber\\
& & \iff \dfrac{(\sqrt{n}\beta_i + U_i)^2}{\log n} < \dfrac{1}{\log n}\dfrac{n+\lambda}{n}RSS_\lambda(\gamma)\left(1-a_n^{\frac{1}{n-1}}\right)
\end{eqnarray}
where $a_n = \lambda w^2/((n+\lambda)(1-w)^2).$ But, $RSS_\lambda(\gamma) \leq \|\tilde{y}\|^2 = n\|\beta_t\|^2 + 2\beta^\top_tX_t^\top \epsilon + \|\epsilon\|^2$ and hence $\|\tilde{y}\|^2/n \to \|\beta_t\|^2 + 1$ almost surely. Recall that as in Theorem~\ref{thm:orthogonal}, $\lambda$ is assumed fixed here. Also, $\log a_n/\log n \to -(2c+1)$ as $n\to \infty,$
\[\dfrac{1}{\log n}\dfrac{n+\lambda}{n}RSS_\lambda(\gamma)\left(1-a_n^{\frac{1}{n-1}}\right) \leq \dfrac{\|\tilde{y}\|^2}{n-1}\dfrac{n+\lambda}{n}\dfrac{\log a_n}{\log n}\dfrac{(1-e^{(\log a_n)/(n-1)})} {(\log a_n)/(n-1)}\to (2c+1)(\|\beta_t\|^2+1),\]
in probability. Hence, $P(A_0) \to 1,$ as $n\to\infty$ where
$$A_0 = \left\{\max_{\gamma\subseteq t}\dfrac{1}{\log n}\dfrac{n+\lambda}{n}RSS_\lambda(\gamma)\left(1-a_n^{\frac{1}{n-1}}\right) < c''\right\}$$ 
for some constant $c'' >0$. Next denote $A = \cap_{i\in t}A_i$ where
$A_i$'s are defined in the proof of Theorem 1. Then
$P(A)\to 1,$ as $n\to\infty$ and $\gamma^{(j)}\subset t$ for
$1\le j \le |t|$ on $A$.

Since $\sqrt{n}\beta_+ \succ \sqrt{\log n},$ we have
$\sqrt{n}|\beta_i| - \sqrt{c''\log n} > \sqrt{2\log n}$ for all $i,$ and for all large
$n$. Hence, from \eqref{proof:maxz1} note that for $1\le j < |t|,$ and
$i \in t\setminus\gamma^{(j)},$ we have, by symmetry of $U_i$
\begin{eqnarray}\label{proof:maxz2}
& & P\left(A_0\cap A \cap \left\{\max_{i\in t\setminus\gamma^{(j)}}f(\gamma^{(j)} + e_i|y) < f(\gamma^{(j)}|y)\right\} \right) \leq P\left( (\sqrt{n}\beta_i + U_i)^2 < c''\log n \right) \nonumber \\
& & \qquad \leq P(U_i > \sqrt{n}|\beta_i|-\sqrt{c''\log n})  \leq P(U_i > \sqrt{2\log n}) \leq \dfrac{1}{n\sqrt{4\pi\log n}}.
\end{eqnarray}
for sufficiently large $n.$ Thus from \eqref{proof:maxz1} and \eqref{proof:maxz2} we have,
\begin{eqnarray*}
P(\mathcal{T} < |t|) & \leq & P\left(A_0\cap A \cap \bigcup_{j=1}^{|t|-1}\left\{\max_{i\in t\setminus \gamma^{(j)}}f(\gamma^{(j)} + e_i|y) < f(\gamma^{(j)}|y)\right\} \right) + P(A_0^c) + P(A^c)\\
& \leq & \sum_{j=1}^{|t|-1}P\left(A_0\cap A \cap \left\{\max_{i\in t\setminus\gamma^{(j)}}f(\gamma^{(j)} + e_i|y) < f(\gamma^{(j)}|y)\right\}\right) + P(A_0^c) + P(A^c) \\
& \leq & |t|/(n\sqrt{4\pi\log n}) + P(A_0^c) + P(A^c),
\end{eqnarray*}
so that $\lim P(\mathcal{T} < |t|) = 0.$

Next, we show that $P(\mathcal{T} \ge |t|+1) \to 0.$ To that end, suppose $j \notin t$. Then the probability of stopping the iteration exactly at
$|t|$ is $P(\max_{j\notin t}f(t + e_j |y) < f(t|y)).$ Since for any $j\notin t, (X_j^\top \tilde{y})^2 = nU_j^2$ where $U_l$'s are i.i.d $N(0,1)$ variables, $\max_{j\notin t}f(t + e_j |y) < f(t|y)$ is equivalent to
\begin{eqnarray}
\label{proof:maxz}
& & \log a_n + (n-1)\log(RSS_\lambda(t)) < (n-1)\min_{j\notin t}\left\{\log\left(RSS_\lambda(t) - {(X_j^\top \tilde{y})^2}/{(n+\lambda)}\right)\right\}\nonumber\\
&& \iff \max_{j\notin t}\dfrac{U_j^2}{\log n} < \dfrac{1}{\log n}\dfrac{n+\lambda}{n}RSS_\lambda(t)\left(1-a_n^{\frac{1}{n-1}}\right),
\end{eqnarray}
where $a_n = \lambda w^2/((n+\lambda)(1-w)^2)$ is as defined before.
We will now show that the left side is less than 2 with probability
tending to 1 and the right side converges to $2c+1 > 2$ in
probability. This will complete the proof.

First, as $n\to\infty,$ we have 
\[P(\max_{j\notin t}U_j^2 > 2\log n) \le \sum_{j\notin t}P(U_j^2 > 2\log n) \le 2n\left(1 - \Phi(\sqrt{2\log n})\right) \leq \dfrac{1}{\sqrt{\pi\log n}}\to 0.\]

Then, note that under the orthogonal design, for any $\gamma \subseteq t,$
\[RSS_\lambda(t) = RSS(t) + \dfrac{\lambda}{n(n+\lambda)}{\tilde{y}^{\top}X_{t}X_{t}^{\top}\tilde{y}},\]
and $RSS(t)/n \to 1,$ in probability, so that $RSS_\lambda(t)/(n-1) \to 1$ in probability. Consequently, as $n\to \infty,$ with probability tending to one,
\[\dfrac{1}{\log n}\dfrac{n+\lambda}{n}RSS_\lambda(t)\left(1-a_n^{\frac{1}{n-1}}\right) = \dfrac{n+\lambda}{n}\dfrac{RSS_\lambda(t)}{n-1} \dfrac{\left(1 - e^{(\log a_n)/(n-1)}\right)}{(\log a_n)/(n-1)}\dfrac{\log a_n}{\log n} ~\to~ 2c+1 > 2.\]
\end{proof}
\begin{proof}[Proof of Theorem \ref{thm.gen_stop}]
Let
\[ B_i = \left\{\max_{k\notin \gamma^{(i)}} (\Omega(\gamma^{(i)}) - \Omega(\gamma^{(i)}+e_k)) < 0 \right\} \cap \{\gamma^{(i)}\nsupseteq t\}.\]
Thus BITS is stopped \emph{prematurely} by the posterior probability criterion  without including all variables in $t$ iff $B_i$ happens for some $i.$ However, note that
\[\cup B_i = \left\{\cup_{i=1}^{K_n|t|} B_i\right\} \cup  \left\{\cup_{i>K_n|t|} B_i\right\}.\]
We first analyze $\cup_{i=1}^{K_n|t|} B_i.$ To that end, note that when $\gamma^{(i)}$ does not contain $t,$ 
\[\max_{k\in t\backslash \gamma^{(i)}} (\Omega(\gamma^{(i)}) - \Omega(\gamma^{(i)}+e_k)) \leq \max_{k\notin \gamma^{(i)}} (\Omega(\gamma^{(i)}) - \Omega(\gamma^{(i)}+e_k)).\]
Consequently, for $i \leq K_n|t|,$ we have
\begin{eqnarray*}
& & B_i \cap \{\sigma\Upsilon_n < s_n\} \subseteq \left\{\max_{k\in t\backslash \gamma^{(i)}} (\Omega(\gamma^{(i)}) - \Omega(\gamma^{(i)}+e_k)) < 0\right\}\cap \{\sigma\Upsilon_n < s_n\} \\
& \subseteq & \left\{\frac{1}{2\|\tilde{y}\|^2}\max_{k\in t\backslash\gamma^{(i)}}(\rss(\gamma^{(i)} - \rss(\gamma^{(i)}+e_k))) < \frac{\lambda}{2n\tau_+} + \frac{\log(n+\lambda)-\log\lambda}{2(n-1)} - \dfrac{\log\frac{w}{1-w}}{n-1} \right\} \cap \\ & \qquad & \{\sigma\Upsilon_n < s_n\}\\
&\subseteq & \left\{\frac{1}{2\|\tilde{y}\|^2} |s_n - \sigma\Upsilon_n|^2 < \frac{c''\log n}{K_n} - \frac{1}{n-1}\log\frac{w}{1-w}   \right\} \cap \{\sigma\Upsilon_n < s_n\}
\end{eqnarray*}
which is independent of $i.$ In the second set inequality above we have used Lemma~\ref{lem.omegalowerbound} and the third inequality is due to Corollary \ref{cor.maxmaxbound} and \eqref{eq.cdoubleprime}. Also, note that 
\[c''\log n-\dfrac{K_n}{n-1}\log\dfrac{w}{1-w} < c''' \log n\]
by condition C3 for some $c'''>0$ for sufficiently large $n.$ Hence, for sufficiently large $n,$
\begin{eqnarray}\label{eq.firstunionB}
\cup_{i=1}^{K_n|t|} B_i & \subseteq & \left\{\{\sigma \Upsilon_n <s_n \}\cap \cup_{i=1}^{K_n|t|} B_i\right\} \cup \{\sigma \Upsilon_n \ge s_n \} \nonumber\\
& \subseteq & \left\{\frac{K_n}{2\|\tilde{y}\|^2} |s_n - \sigma \Upsilon_n|^2 < c'''\log n, ~\sigma\Upsilon_n < s_n   \right\} \cup \{\sigma\Upsilon_n \ge s_n\} \nonumber \\ 
 & \subseteq & \left\{
 s_n\left(1 - \sqrt{{\|\tilde{y}\|^2}/({nu_n'})}\right)
\le \sigma \Upsilon_n  \right\}
\end{eqnarray}
where $u_n' = \tau_+^2\beta_+^4K_n/(2c'''|t|\|\beta_t\|^2\log n).$ Note from the \eqref{eq.finalbound} that
\begin{equation}\label{eq.secondunionB}
\cup_{i>K_n|t|} B_i \subseteq \{\theta_{|t|} > K_n|t|\} \subseteq \left\{
 s_n\left(1 - \sqrt{{\|\tilde{y}\|^2}/({nu_n})}\right)
\le \sigma \Upsilon_n  \right\}. 
\end{equation}
Now let $c^* = \max\{c,2c'''\}$ and $u^*_n=\min\{u_n,u_n'\}= \tau_+^2\beta_+^4K_n/(c^*|t|\|\beta_t\|^2\log n).$   Then, for all sufficiently large $n,$ combining \eqref{eq.firstunionB} and \eqref{eq.secondunionB} we finally get,
\begin{eqnarray*}
P(\cup B_i) & \le & P\left(
\sigma \Upsilon_n \geq s_n\left(1 - \sqrt{{\|\tilde{y}\|^2}/({nu_n^*})}\right)
 \right) \\
&\le & \exp(1 - q(s_n\delta) + K_n|t|\log p + \log|t|) +  P(\|\tilde{y}\|^2 > nu_n^* (1-\sigma \delta)^2).
\end{eqnarray*}
\end{proof}

\section{Proofs of lemmas and corollaries}\label{sec:supp_proofs}
\subsection{Proof of Lemma \ref{lem:Kexist}}\label{sec:Kexist}
\begin{proof}
  Since $K_n = n^{\xi_0+4\xi_{min}} (\log n)^2$,
  \[
    \dfrac{|t|\|\beta_t\|^2\log n }{\tau_{+}^2\beta_{+}^4} \le \nu \nu_{\beta}^{-4} \tau_{min}^{-2} C_{\beta}^2 n^{\xi_0 + 4\xi_{min}} \log n  \prec K_n.
    \]
    Also, since $\lambda$ and $w$ are fixed here, and $\tau_+\geq \tau_{{min}}$,
    \[
      n\log n \min\left\{\frac{\tau_+}{\lambda}, \frac{1}{|\log (1/w-1)|} \right\} \ge c_* n\log n
    \]
    for some constant $c_*$. Thus, if $\xi_0 + 4\xi_{min} <1$,
\[
    \dfrac{|t|\|\beta_t\|^2\log n }{\tau_{+}^2\beta_{+}^4} \prec K_n \preceq n\log n \min\left\{\frac{\tau_+}{\lambda}, \frac{1}{|\log (1/w-1)|} \right\}. \]

  Next, since $\epsilon_i \stackrel{iid}{\sim}{\cal N}(0,1)$, $q(\zeta) = \zeta^2/2$. Therefore,
  \begin{align}
    \label{eq:qsnlb}
    q(s_n\delta) - K_n|t|\log p - \log|t| &\ge \frac{n\tau^2_+\beta_+^4 \delta^2}{2\|\beta_t\|^2|t|} - K_n|t|\log p - \log|t|\nonumber\\
    &\ge 0.5 \delta^2 \nu^{-1} \nu_{\beta}^{4} \tau_{min}^{2} C_{\beta}^{-2} n^{1- \xi_0 - 4 \xi_{min}} - \nu^2 n^{\xi+2\xi_0+4\xi_{min}} (\log n)^2 \nonumber\\ & \hspace{.1in} - \nu \xi_0 \log n 
  \end{align}
 
Note that \eqref{eq:qsnlb} $\rightarrow \infty$  because $\xi + 3\xi_0 + 8\xi_{min}<1$.
\end{proof}

\subsection{Proof of Lemma \ref{lemma:fw1}}
\label{sec:prooflemmafw1}

\begin{proof}
Note that $RSS_\lambda(\gamma) = nv_{y\cdot\gamma,\lambda}. $
Suppose ${U}_\gamma$ denotes the upper triangular Cholesky factor of ${X}_\gamma^\top{X}_\gamma + \lambda{I}.$ And let for $i\notin \gamma,$ $\gamma' = \gamma + e_i.$ Then arranging the columns of ${X}_{\gamma'}$ appropriately, we can assume that the Cholesky factor of ${X}_{\gamma'}^\top{X}_{\gamma'} + \lambda{I}$ is given by
\[{U}_{\gamma'} = \begin{pmatrix}
                          {U}_\gamma & {s} \\ {0} & s_0
                         \end{pmatrix}
\textrm{ so that }
{U}_{\gamma'}^{-\top}{X}_{\gamma'}^\top\tilde{{y}} =
\begin{pmatrix}
{U}_\gamma^{-\top}{X}_\gamma\tilde{{y}}\\
\left({X}_i^\top \tilde{{y}} - {s}^\top {U}_\gamma^{-\top}{X}_\gamma\tilde{{y}}\right)/s_0
\end{pmatrix}
= \begin{pmatrix}
{U}_\gamma^{-\top}{X}_\gamma\tilde{{y}}\\
\sqrt{n}v_{iy\cdot\gamma,\lambda}/\sqrt{v_{i\cdot\gamma,\lambda} }
\end{pmatrix}
\]
where ${s} = {U}_\gamma^{-\top}{X}_\gamma^\top{X}_i$ and $s_0^2 = {X}_i^\top{X}_i + \lambda - {s}^\top{s} = nv_{i\cdot\gamma,\lambda}.$ Also note that,
\begin{eqnarray*}
 RSS_\lambda(\gamma) - RSS_\lambda(\gamma') & = &\| {U}_{\gamma'}^{-\top}{X}_{\gamma'}^\top\tilde{{y}} \|^2_2 - \| {U}_{\gamma}^{-\top}{X}_{\gamma}^\top\tilde{{y}} \|^2_2 = n\{v_{iy\cdot\gamma,\lambda}\}^2/v_{i.\gamma,\lambda} \\
 \Rightarrow \{RSS_\lambda(\gamma) - RSS_\lambda(\gamma')\}/RSS_\lambda(\gamma) & = & R^2_{i\cdot\gamma,\lambda} \\
 \Rightarrow \log RSS_\lambda(\gamma') - \log RSS_\lambda(\gamma) & = & \log\left(1-R^2_{i\cdot\gamma,\lambda}\right).
\end{eqnarray*}
Also,
\[\log|{X}_{\gamma'}^\top{X}_{\gamma'} + \lambda{I}| - \log|{X}_{\gamma}^\top{X}_{\gamma} + \lambda{I}| = 2\log s_0 = \log \left(nv_{i\cdot\gamma,\lambda}.\right)\]

Therefore, \begin{eqnarray*}
\log f(\gamma'|y) - \log f(\gamma |y) &=& \dfrac{1}{2}\log \lambda - \dfrac{1}{2} \left(\log|{X}_{\gamma'}^\top{X}_{\gamma'} + \lambda{I}| - \log|{X}_{\gamma}^\top{X}_{\gamma} + \lambda{I}|\right)\\
&-& \dfrac{n-1}{2} \left(\log RSS_\lambda(\gamma') - \log RSS_\lambda(\gamma)\right) + \log(w/(1-w))\\
&=& \half\log(n\lambda w^2/(1-w)^2) - \half\log v_{i\cdot\gamma,\lambda} - \half(n-1)\log \{1 - R^2_{i\cdot\gamma,\lambda}, \}
\end{eqnarray*}

which completes the proof.
\end{proof}

\subsection{Proof of Lemma \ref{lem.rssdiff}}
\begin{proof}
  Since
  \[
    \rss(\gamma) - \rss(\gamma+e_i) =  (\tilde{y}^{\top}X_\gamma \; \tilde{y}^{\top}X_i) \begin{pmatrix}
X_\gamma^\top X_\gamma &  X_\gamma^\top X_i\\
X_i^\top X_\gamma & n
\end{pmatrix}^{-1} \begin{pmatrix}
 X_\gamma^\top \tilde{y}\\
X_i^\top \tilde{y}
\end{pmatrix} -\tilde{y}^{\top}X_\gamma (X_\gamma^\top X_\gamma)^{-1}X_\gamma^\top \tilde{y},
\]
the proof follows from the fact that
\begin{align*}
&  \begin{pmatrix}
X_\gamma^\top X_\gamma &  X_\gamma^\top X_i\\
X_i^\top X_\gamma & n
\end{pmatrix}^{-1}\\ & = \frac{1}{a}   \begin{pmatrix}
a(X_\gamma^\top X_\gamma)^{-1} + (X_\gamma^\top X_\gamma)^{-1}  X_\gamma^\top X_i X_i^\top X_\gamma (X_\gamma^\top X_\gamma)^{-1} &  -(X_\gamma^\top X_\gamma)^{-1}X_\gamma^\top X_i\\
-X_i^\top X_\gamma(X_\gamma^\top X_\gamma)^{-1}  & 1
\end{pmatrix},
\end{align*}

where $a=n - X_i^\top X_\gamma^\top(X_\gamma^\top X_\gamma)^{-1}X_\gamma^\top X_i$.
\end{proof}

\subsection{Proof of Lemma \ref{lem.logdetdiffbound}}
\begin{proof}
 Note that,
 \[\log|X_{\gamma'}^\top X_{\gamma'} + \lambda I| - \log|X_{\gamma}X_{\gamma}+\lambda I| = \log(n + \lambda - X_j^\top X_\gamma (X_\gamma^\top X_\gamma + \lambda I)^{-1}X_\gamma^\top X_j),\]
 because $X_j^\top X_j = n.$ Since $X_j^\top X_\gamma (X_\gamma^\top X_\gamma + \lambda I)^{-1}X_\gamma^\top X_j) \ge 0$ (equality holding iff $X_\gamma^\top X_j = 0)$ and $n - X_j^\top X_\gamma (X_\gamma^\top X_\gamma + \lambda I)^{-1}X_\gamma^\top X_j) > 0,$ the result follows immediate.
\end{proof}

\subsection{Proof of Lemma \ref{lem.constants}}
\begin{proof}
 To prove \eqref{eq.cprime} note that
 \[ \log(1+nK_n|t|/\lambda) \le \max\{\log 2, \log(2nK_n|t|/\lambda) = O(\log n), \]
 since $K_n|t| < n$ and $|\log\lambda| = O(\log n).$ To prove \eqref{eq.cdoubleprime} note that $K_n\lambda/(n\tau_+) = O(\log n)$ and $K_n< n.$
 
\end{proof}

\subsection{Proof of Lemma \ref{lem.omegalowerbound}}
\begin{proof}
Note that for any $k \notin \gamma,$
\begin{eqnarray*}\label{eq.omegalowerbound}
& & \Omega(\gamma) - \Omega(\gamma+e_k) \nonumber \\
& \ge & \frac12 \left(\log\rssl(\gamma) - \log\rssl(\gamma+e_k)\right) - \dfrac{\log(n+\lambda)}{2(n-1)} + \dfrac{\log\lambda}{2(n-1)} + \frac{1}{n-1}\log\dfrac{w}{1-w} \nonumber\\
& = & \frac12 \left(\log\dfrac{\rssl(\gamma)}{\|\tilde{y}\|^2} - \log\dfrac{\rssl(\gamma+e_k)}{\|\tilde{y}\|^2}\right) - \dfrac{\log(n+\lambda)}{2(n-1)} + \dfrac{\log\lambda}{2(n-1)} + \frac{1}{n-1}\log\dfrac{w}{1-w} \nonumber\\
& \ge & \dfrac{1}{2\|\tilde{y}\|^2}\left(\rssl(\gamma)-\rssl(\gamma+e_k)\right) - \dfrac{\log(n+\lambda)}{2(n-1)} + \dfrac{\log\lambda}{2(n-1)} + \frac{1}{n-1}\log\dfrac{w}{1-w} \nonumber\\
& = & \dfrac{1}{2\|\tilde{y}\|^2}\left(\{\rssl(\gamma) - \rss(\gamma)\} + \{\rss(\gamma+e_k) - \rssl(\gamma+e_k)\} \right. + \nonumber \\
 & & \left. \rss(\gamma) - \rss(\gamma+e_k)\right) - \dfrac{\log(n+\lambda)}{2(n-1)} + \dfrac{\log\lambda}{2(n-1)} + \frac{1}{n-1}\log\dfrac{w}{1-w} \nonumber\\
 & \ge & \dfrac{1}{2\|\tilde{y}\|^2}\left(\{0\} - \dfrac{\lambda\|\tilde{y}\|^2}{n\tau_+} + \rss(\gamma) - \rss(\gamma+e_k)\right) - \dfrac{\log(n+\lambda)}{2(n-1)} + \dfrac{\log\lambda}{2(n-1)} \nonumber\\ & &\hspace*{.1in}+ \frac{1}{n-1}\log\dfrac{w}{1-w} \nonumber\\
 & = & \dfrac{1}{2\|\tilde{y}\|^2}\left(\rss(\gamma) - \rss(\gamma+e_k)\right) - \dfrac{\lambda}{2n\tau_+}- \dfrac{\log(n+\lambda)}{2(n-1)} + \dfrac{\log\lambda}{2(n-1)} + \frac{1}{n-1}\log\dfrac{w}{1-w}.
\end{eqnarray*}
In the above, the first inequality follows from the definition of $\Omega(\gamma)$ and Lemma \ref{lem.logdetdiffbound}; the second inequality follows from the facts that $\rssl(\gamma) < \|\tilde{y}\|^2$ for any model $\gamma$ and that $\log u - \log v \ge u-v$ for $0<v<u<1.$ Finally, the last inequality follows from the facts that $\rssl(\gamma) \ge \rss(\gamma)$ for any $\gamma$ and that for all $\gamma'$ of size at most $(K_n+1)|t|$ we have
\begin{eqnarray*}
\rssl(\gamma') - \rss(\gamma') & = & \tilde{y}^\top X_{\gamma'} \bigg( (X_{\gamma'}^\top X_{\gamma'})^{-1} - (X_{\gamma'}^\top X_{\gamma'} + \lambda I)^{-1} \bigg)X_{\gamma'}^\top \tilde{y} \\
& = & \tilde{y}^\top X_{\gamma'} (X_{\gamma'}^\top X_{\gamma'})^{-1/2}\bigg( I - (I + \lambda (X_{\gamma'}^\top X_{\gamma'})^{-1})^{-1}\bigg)(X_{\gamma'}^\top X_{\gamma'})^{-1/2}X_{\gamma'}^\top \tilde{y} \\
&\leq & \lambda \tilde{y}^\top X_{\gamma'} (X_{\gamma'}^\top X_{\gamma'})^{-1/2} (X_{\gamma'}^\top X_{\gamma'})^{-1}(X_{\gamma'}^\top X_{\gamma'})^{-1/2}X_{\gamma'}^\top \tilde{y} \\
&\leq & \lambda \tilde{y}^\top X_{\gamma'} (X_{\gamma'}^\top X_{\gamma'})^{-2}X_{\gamma'}^\top \tilde{y} \le \dfrac{\lambda \|\tilde{y}\|^2}{n\tau_+},
\end{eqnarray*}
because the nonzero eigenvalues of $X_{\gamma'} (X_{\gamma'}^\top X_{\gamma'})^{-2}X_{\gamma'}^\top$ and $ (X_{\gamma'}^\top X_{\gamma'})^{-1}$ are the same.
\end{proof}

\subsection{Proof of Lemma \ref{lem.regssbound}}
\begin{proof}
Note that
\begin{eqnarray*} \label{eq.regssbound}
& & \|\beta_t\|^2 |t| \max_{k\in t\backslash\gamma} (X_k^\top (I-H_\gamma)X_t\beta_t)^2 \geq \|\beta_t\|^2 \sum_{k\in t\backslash\gamma} (X_k^\top (I-H_\gamma)X_t\beta_t)^2 \nonumber\\
& = & \|\beta_t\|^2 \sum_{k\in t} (X_k^\top (I-H_\gamma)X_t\beta_t)^2 \nonumber\\
& \geq & \left( \sum_{k\in t} \beta_k X_k^\top (I-H_\gamma)X_t\beta_t \right)^2 \nonumber\\
& = & \{\beta_t^\top X_t^\top (I-H_\gamma)X_t\beta_t\}^2 \nonumber\\
& = & \| (I - H_\gamma)X_t\beta_t \|^4  = \| X_t\beta_t - X_{\gamma} (X_{\gamma}^\top X_{\gamma})^{-1}X_{\gamma}X_t\beta_t \|^4 = \| X_{t\backslash\gamma}\beta_{t\backslash\gamma} + X_{\gamma}\tilde{b}\|^4 \nonumber\\
& \geq & n^2\tau_+^2\beta_+^4,
\end{eqnarray*}
for some vector $\tilde{b},$ where the first equality follows from the fact that $(I-H_\gamma)X_k = 0$ for any $k\in t\cap \gamma,$ the second inequality is the Cauch-Schwarz inequality, and the final inequality follows from the fact that the matrix $[X_{t\backslash\gamma} \quad X_{\gamma}]$ has less than $(K_n+1)|t|$ columns and that $\|\beta_{t\backslash\gamma}\| \geq \beta_+.$
The proof follows because $s_n = \sqrt{n}\tau_+\beta_+^2/(\|\beta_t\|\sqrt{|t|}).$
\end{proof}

\subsection{Proof of Corollary \ref{cor:consistency}}
\begin{proof}
  Since $\tilde{y}^{\top}\tilde{y} \le \beta_{t}^{\top}X_{t}^{\top}X_{t}\beta_{t}+\sigma^2{\epsilon}^{\top}{\epsilon}+2\sigma {\epsilon}^{\top}X_{t}\beta_{t}$, 
  we have
\begin{eqnarray}
  \label{eq:ytilin}
P(\|\tilde{y}\|^2 > n u_n) &\le& P\left( \frac{\beta_{t}^{\top}X_{t}^{\top}X_{t}\beta_{t}}{n} + \frac{\sigma^2{\epsilon}^{\top}{\epsilon}}{n}+\frac{2\sigma {\epsilon}^{\top}X_{t}\beta_{t}}{n} >  u_n\right) \nonumber\\
&\leq& P\left(\frac{\sigma^2{\epsilon}^{\top}{\epsilon}}{n} > v_n \right) + P\left(\frac{2\sigma {\epsilon}^{\top}X_{t}\beta_{t}}{n} > 1\right).
\end{eqnarray}

By Berry-Esseen theorem, 
\begin{eqnarray*}
P\left(\frac{\sigma^2{\epsilon}^{\top}{\epsilon}}{n} > v_n\right) &\leq 1 - \Phi\left(\dfrac{\sqrt{n}v_n}{\sigma^2 \sqrt{\kappa}}\right) + \dfrac{c_1}{\sqrt{n}}\le \frac{\sigma^2 \sqrt{\kappa}}{\sqrt{2n\pi} v_n}\exp\left(-\dfrac{nv_n^2}{2\sigma^4\kappa}\right) + \dfrac{c_1}{\sqrt{n}},
\end{eqnarray*}
where $v_n >0$ for all large $n$. 
 Since Var $({\epsilon}^{\top}X_{t}\beta_{t}) = \sigma^2 \beta_t^\top X_t^\top X_{t}\beta_{t}$, the
proof follows as by the Chebyshev's inequality we have
\begin{equation*}\label{eq:lemma4bound2}
    \begin{split}
        P\left(\frac{2\sigma {\epsilon}^{\top}X_{t}\beta_{t}}{n} > 1\right) \leq \frac{4\sigma^4\beta_{t}^{\top}X_{t}^{\top}X_{t}\beta_{t}}{n^2} 
    \end{split}
  \end{equation*}
\end{proof}

\subsection{Proof of Corollary \ref{cor.maxmaxbound}}
\begin{proof}
Let $\tilde{\epsilon} = (I_n - n^{-1}1_n1_n^\top)\epsilon.$ Note that $H_\gamma 1_n = 0$ since $X^\top 1_n = 0.$ Thus, 
\[X_k^\top (I-H_\gamma)\tilde\epsilon = X_k^\top (I-H_\gamma)\epsilon.\]
Consequently, from Lemma \ref{lem.rssdiff} we have 
\begin{eqnarray*}\label{eq.rssdiff2}
& & \max_{k\in t\backslash\gamma} \{\rss(\gamma) - \rss(\gamma+e_k)\} = \max_{k\in t\backslash\gamma}\dfrac{(X_k^\top(I-H_\gamma)\tilde{y})^2}{\| (I-H_\gamma)X_k\|^2} \nonumber \\
& = & \max_{k\in t\backslash\gamma}\dfrac{\left\{X_k^\top(I-H_\gamma)X_t\beta_t + \sigma X_k^\top(I-H_\gamma)\epsilon\right\}^2}{\| (I-H_\gamma)X_k\|^2} \nonumber\\
& \geq & \left|\max_{k\in t\backslash\gamma}\dfrac{|X_k^\top(I-H_\gamma)X_t\beta_t|}{\| (I-H_\gamma)X_k\|} -  \sigma \max_{k\in t\backslash\gamma}\dfrac{|X_k^\top(I-H_\gamma)\epsilon|}{\| (I-H_\gamma)X_k\|} \right|^2 \nonumber\\
& \geq & \left|\dfrac{n\tau_+\beta_+^2\|\beta_t\|^{-1}|t|^{-1/2}}{\sqrt{n}} -  \sigma \max_{k\in t}\max_{\gamma:|\gamma|<K_n|t|} |a_{\gamma,l}^\top \epsilon| \right|^2  \nonumber\\
& = & \left|s_n -  \sigma \Upsilon_n \right|^2.
\end{eqnarray*}
In the above, the first inequality follows from the fact that for any two sequences $(b_m)$ and $(c_m)$ of real numbers,
$\max_m |b_m - c_m| \geq |\max_m |b_m| - \max_m |c_m||;$ the last inequality follows from Lemma \ref{lem.regssbound} and from the assumption that $\sigma\Upsilon_n < s_n.$
\end{proof}


\section{Further simulation results}\label{sec:simufurther}


In this section we provide simulation results from $(n,p) = (75,200),$ 
$(100,2000),$ $(150, 2000)$ The eight simulation models we consider here are 
described in Section~\ref{sec:exam}.

\begin{table}[ht]
\centering
\caption{n = 75, p = 200} 
\begin{tabular}{lrrrrrrrr}
  \hline
Method & Ind. & CS & AR & Fac. & Grp. & Ext. & Sp.Fac. & Spur. \\ 
  \hline
\multicolumn{9}{l}{Mean true positive rates}\\
BITS1(n) & 86.4 & 81.6 & 87.6 & 84.6 & 90.7 & 79.8 & 85.8 & 6.4 \\ 
  BITS1(PP) & 76.7 & 69.7 & 71.0 & 69.7 & 77.7 & 78.1 & 62.3 & 5.4 \\ 
  BITS2(n) & 86.0 & 80.8 & 86.6 & 86.0 & 90.1 & 79.8 & 86.4 & 5.6 \\ 
  BITS2(PP) & 76.4 & 69.1 & 71.7 & 72.6 & 74.8 & 77.6 & 62.4 & 4.0 \\ 
  BITS3(n) & 86.0 & 81.4 & 85.8 & 87.8 & 92.2 & 79.7 & 84.7 & 4.2 \\ 
  BITS3(PP) & 77.1 & 70.2 & 69.3 & 70.2 & 49.6 & 77.2 & 64.0 & 2.9 \\ 
  UBITS(n) & 90.9 & 87.7 & 92.7 & 90.8 & 95.3 & 82.0 & 92.4 & 7.6 \\ 
  UBITS(PP) & 79.3 & 75.3 & 75.1 & 80.4 & 81.6 & 79.4 & 72.0 & 6.1 \\ 
  HOLP(n) & 79.3 & 78.4 & 81.2 & 79.6 & 82.4 & 84.7 & 70.2 & 7.9 \\ 
  HOLP(eBIC) & 40.1 & 33.1 & 36.9 & 48.0 & 35.8 & 50.1 & 13.7 & 0.1 \\ 
  SIS(n) & 76.7 & 60.7 & 81.1 & 55.6 & 87.3 & 53.1 & 69.4 & 0.0 \\ 
  SIS(eBIC) & 34.2 & 19.1 & 30.0 & 11.8 & 37.2 & 34.1 & 9.4 & 0.0 \\ 
  FR(n-2) & 84.6 & 77.8 & 79.6 & 86.3 & 51.9 & 87.0 & 70.8 & 35.8 \\ 
  FR(eBIC) & 60.7 & 46.4 & 48.3 & 56.8 & 25.9 & 58.0 & 30.1 & 0.0 \\ 
   \hline
\multicolumn{9}{l}{Mean posterior mass coverages}\\
BITS1(n) & 69.3 & 66.8 & 69.3 & 53.6 & 78.0 & 67.1 & 52.2 & 67.1 \\ 
  BITS1(PP) & 64.5 & 60.3 & 63.9 & 47.9 & 69.4 & 60.3 & 41.6 & 52.2 \\ 
  BITS2(n) & 70.2 & 66.6 & 69.1 & 55.4 & 77.1 & 66.4 & 55.5 & 66.0 \\ 
  BITS2(PP) & 65.0 & 60.3 & 63.6 & 49.0 & 68.6 & 59.6 & 42.5 & 51.3 \\ 
  BITS3(n) & 69.5 & 64.7 & 67.8 & 55.5 & 75.4 & 64.0 & 54.3 & 60.7 \\ 
  BITS3(PP) & 64.3 & 58.3 & 62.5 & 49.8 & 67.8 & 57.8 & 42.0 & 48.0 \\ 
  UBITS(n) & 92.4 & 92.6 & 92.6 & 86.4 & 98.2 & 88.2 & 95.2 & 98.2 \\ 
  UBITS(PP) & 87.1 & 86.2 & 87.0 & 80.5 & 91.6 & 82.0 & 83.8 & 86.3 \\ 
   \hline
\multicolumn{9}{l}{Median model sizes}\\
BITS1(PP) & 17.5 & 23.0 & 18.0 & 17.0 & 28.0 & 20.0 & 22.0 & 42.0 \\ 
  BITS2(PP) & 19.0 & 24.0 & 19.0 & 16.0 & 28.0 & 20.0 & 23.0 & 39.0 \\ 
  BITS3(PP) & 19.5 & 26.5 & 16.0 & 15.0 & 13.5 & 9.0 & 26.0 & 34.0 \\ 
  UBITS(n) & 109.0 & 108.0 & 109.0 & 101.0 & 109.0 & 99.0 & 107.0 & 93.0 \\ 
  UBITS(PP) & 27.0 & 40.0 & 25.0 & 25.0 & 41.0 & 32.0 & 33.0 & 55.0 \\ 
  HOLP(eBIC) & 3.0 & 3.0 & 3.0 & 6.0 & 4.0 & 5.0 & 2.0 & 1.0 \\ 
  SIS(eBIC) & 3.0 & 2.0 & 2.0 & 3.0 & 4.0 & 3.0 & 2.0 & 1.0 \\ 
  FR(eBIC) & 6.0 & 4.0 & 5.0 & 9.0 & 2.0 & 5.0 & 6.0 & 1.0 \\ 
   \hline
\multicolumn{9}{l}{}\\
\end{tabular}
\label{tab.n75p200}
\end{table}

\begin{table}[ht]
\centering
\caption{n = 100, p = 2000} 
\begin{tabular}{lrrrrrrrr}
  \hline
Method & Ind. & CS & AR & Fac. & Grp. & Ext. & Sp.Fac. & Spur. \\ 
  \hline
\multicolumn{9}{l}{Mean true positive rates}\\
BITS1(n) & 70.3 & 62.0 & 75.3 & 56.7 & 82.4 & 80.4 & 57.6 & 7.6 \\ 
  BITS1(PP) & 69.3 & 61.6 & 69.3 & 56.7 & 80.9 & 80.4 & 50.7 & 7.6 \\ 
  BITS2(n) & 71.4 & 62.2 & 72.4 & 66.0 & 82.0 & 77.6 & 59.3 & 0.3 \\ 
  BITS2(PP) & 70.6 & 60.0 & 62.3 & 59.0 & 59.2 & 77.2 & 44.1 & 0.3 \\ 
  BITS3(n) & 71.7 & 61.1 & 70.6 & 65.3 & 82.0 & 77.4 & 59.1 & 0.1 \\ 
  BITS3(PP) & 70.9 & 59.3 & 62.4 & 55.3 & 41.7 & 76.8 & 43.8 & 0.1 \\ 
  UBITS(n) & 75.2 & 68.9 & 82.7 & 71.2 & 84.8 & 83.8 & 70.0 & 7.7 \\ 
  UBITS(PP) & 73.9 & 68.1 & 76.1 & 72.2 & 82.2 & 83.4 & 60.8 & 7.7 \\ 
  HOLP(n) & 59.0 & 57.3 & 67.0 & 53.2 & 75.9 & 85.1 & 47.1 & 32.9 \\ 
  HOLP(eBIC) & 32.8 & 25.8 & 27.8 & 30.4 & 31.9 & 56.0 & 11.0 & 9.6 \\ 
  SIS(n) & 58.4 & 40.4 & 67.8 & 20.4 & 75.7 & 53.4 & 47.0 & 0.0 \\ 
  SIS(eBIC) & 32.4 & 17.3 & 27.8 & 4.7 & 31.9 & 38.1 & 9.9 & 0.0 \\ 
  FR(n-2) & 71.0 & 61.0 & 63.1 & 59.2 & 31.8 & 79.1 & 44.0 & 4.2 \\ 
  FR(eBIC) & 60.2 & 44.2 & 48.9 & 36.2 & 25.4 & 57.6 & 28.5 & 0.0 \\ 
   \hline
\multicolumn{9}{l}{Mean posterior mass coverages}\\
BITS1(n) & 53.6 & 44.3 & 51.2 & 10.7 & 80.7 & 66.1 & 12.1 & 73.4 \\ 
  BITS1(PP) & 52.6 & 42.9 & 50.1 & 8.8 & 76.9 & 63.5 & 11.1 & 61.7 \\ 
  BITS2(n) & 64.6 & 54.8 & 62.6 & 22.9 & 76.4 & 64.9 & 39.5 & 46.8 \\ 
  BITS2(PP) & 63.4 & 53.3 & 61.3 & 22.0 & 73.4 & 62.7 & 37.3 & 41.5 \\ 
  BITS3(n) & 64.2 & 54.1 & 60.1 & 26.7 & 75.4 & 61.5 & 37.5 & 33.5 \\ 
  BITS3(PP) & 63.1 & 52.7 & 59.0 & 26.0 & 72.7 & 59.6 & 35.5 & 30.6 \\ 
  UBITS(n) & 77.7 & 75.3 & 77.4 & 49.8 & 91.0 & 83.2 & 62.5 & 93.5 \\ 
  UBITS(PP) & 76.1 & 73.2 & 75.8 & 47.5 & 88.1 & 80.0 & 59.8 & 85.2 \\ 
   \hline
\multicolumn{9}{l}{Median model sizes}\\
BITS1(PP) & 54.0 & 80.5 & 51.0 & 100.0 & 40.0 & 92.0 & 51.5 & 100.0 \\ 
  BITS2(PP) & 38.0 & 39.0 & 39.0 & 36.5 & 44.0 & 36.0 & 37.5 & 48.0 \\ 
  BITS3(PP) & 45.0 & 45.0 & 45.5 & 40.5 & 45.0 & 10.0 & 44.0 & 40.5 \\ 
  UBITS(n) & 186.0 & 186.0 & 186.0 & 182.0 & 183.0 & 181.0 & 182.0 & 173.0 \\ 
  UBITS(PP) & 100.5 & 130.0 & 100.0 & 145.0 & 94.5 & 100.5 & 95.0 & 153.5 \\ 
  HOLP(eBIC) & 3.0 & 2.0 & 2.0 & 5.0 & 1.0 & 5.0 & 2.0 & 2.0 \\ 
  SIS(eBIC) & 3.0 & 2.0 & 2.0 & 3.0 & 1.0 & 3.0 & 2.0 & 1.0 \\ 
  FR(eBIC) & 5.5 & 4.0 & 4.0 & 8.0 & 2.0 & 5.0 & 6.5 & 1.0 \\ 
   \hline
\multicolumn{9}{l}{}\\
\end{tabular}
\label{tab.n100p2000}
\end{table}

\begin{table}[ht]
\centering
\caption{n = 150, p = 2000} 
\begin{tabular}{lrrrrrrrr}
  \hline
Method & Ind. & CS & AR & Fac. & Grp. & Ext. & Sp.Fac. & Spur. \\ 
  \hline
\multicolumn{9}{l}{Mean true positive rates}\\
BITS1(n) & 79.4 & 69.7 & 87.1 & 69.9 & 93.7 & 84.2 & 79.3 & 5.9 \\ 
  BITS1(PP) & 77.0 & 67.6 & 76.2 & 69.8 & 89.3 & 84.2 & 69.0 & 5.9 \\ 
  BITS2(n) & 80.1 & 70.9 & 82.0 & 79.6 & 91.7 & 82.1 & 85.8 & 0.2 \\ 
  BITS2(PP) & 78.9 & 69.2 & 70.4 & 76.8 & 74.4 & 81.7 & 75.6 & 0.2 \\ 
  BITS3(n) & 80.0 & 70.3 & 81.3 & 80.6 & 91.2 & 81.9 & 85.7 & 0.2 \\ 
  BITS3(PP) & 78.9 & 68.8 & 71.0 & 75.9 & 51.0 & 81.4 & 75.6 & 0.2 \\ 
  UBITS(n) & 82.7 & 75.2 & 91.3 & 83.1 & 93.9 & 85.8 & 90.0 & 6.0 \\ 
  UBITS(PP) & 80.2 & 72.9 & 80.7 & 81.4 & 90.3 & 85.8 & 82.0 & 6.0 \\ 
  HOLP(n) & 67.4 & 63.8 & 71.7 & 65.2 & 87.6 & 88.8 & 58.2 & 26.2 \\ 
  HOLP(eBIC) & 42.2 & 37.9 & 40.3 & 43.2 & 44.2 & 64.7 & 18.1 & 4.4 \\ 
  SIS(n) & 67.7 & 46.0 & 71.4 & 24.6 & 87.9 & 56.2 & 58.0 & 0.0 \\ 
  SIS(eBIC) & 41.4 & 22.2 & 38.0 & 5.8 & 43.9 & 43.4 & 17.0 & 0.0 \\ 
  FR(n-2) & 80.1 & 70.0 & 70.6 & 79.1 & 35.2 & 83.9 & 75.0 & 6.8 \\ 
  FR(eBIC) & 69.8 & 56.2 & 58.6 & 65.4 & 28.4 & 66.3 & 62.0 & 0.0 \\ 
   \hline
\multicolumn{9}{l}{Mean posterior mass coverages}\\
BITS1(n) & 69.4 & 62.9 & 69.5 & 16.3 & 84.1 & 76.3 & 26.1 & 76.5 \\ 
  BITS1(PP) & 67.8 & 60.6 & 67.9 & 15.9 & 80.1 & 72.7 & 25.0 & 64.1 \\ 
  BITS2(n) & 73.1 & 65.1 & 73.2 & 38.8 & 80.5 & 73.9 & 62.4 & 59.2 \\ 
  BITS2(PP) & 71.6 & 63.0 & 71.7 & 37.6 & 77.2 & 70.3 & 59.0 & 51.6 \\ 
  BITS3(n) & 72.9 & 64.2 & 71.7 & 40.2 & 78.7 & 70.2 & 62.3 & 44.1 \\ 
  BITS3(PP) & 71.3 & 62.3 & 70.3 & 39.1 & 75.9 & 67.3 & 59.1 & 39.5 \\ 
  UBITS(n) & 86.0 & 83.2 & 86.3 & 62.2 & 94.4 & 91.6 & 77.9 & 97.3 \\ 
  UBITS(PP) & 84.0 & 80.5 & 84.4 & 60.6 & 91.3 & 87.6 & 74.3 & 88.7 \\ 
   \hline
\multicolumn{9}{l}{Median model sizes}\\
BITS1(PP) & 43.0 & 69.0 & 48.0 & 150.0 & 48.5 & 72.5 & 45.0 & 150.0 \\ 
  BITS2(PP) & 56.0 & 56.5 & 56.0 & 44.0 & 65.5 & 10.0 & 45.0 & 74.0 \\ 
  BITS3(PP) & 68.0 & 68.0 & 68.0 & 55.0 & 69.5 & 9.0 & 58.0 & 62.0 \\ 
  UBITS(n) & 276.0 & 274.0 & 276.0 & 269.0 & 273.0 & 265.0 & 268.0 & 250.0 \\ 
  UBITS(PP) & 119.5 & 148.5 & 132.0 & 199.0 & 137.5 & 91.0 & 99.5 & 222.0 \\ 
  HOLP(eBIC) & 4.0 & 3.0 & 3.0 & 8.0 & 4.0 & 6.0 & 4.0 & 2.0 \\ 
  SIS(eBIC) & 4.0 & 2.0 & 3.0 & 3.0 & 4.0 & 4.0 & 3.5 & 1.0 \\ 
  FR(eBIC) & 7.0 & 5.0 & 5.0 & 13.0 & 3.0 & 6.0 & 16.0 & 1.0 \\ 
   \hline
\multicolumn{9}{l}{}\\
\end{tabular}
\label{tab.n150p2000}
\end{table}

\end{document}